\documentclass[aps,prd,onecolumn,groupedaddress,nofootinbib]{revtex4}
\usepackage{graphics}

\begin{document}

\title{Improved Calculation of the Primordial Gravitational Wave Spectrum in the Standard Model}
\author{Yuki Watanabe}
\email[]{yuki@astro.as.utexas.edu}
\affiliation{Department of Physics, University of Texas, Austin, Texas 78712, USA}
\author{Eiichiro Komatsu}
\affiliation{Department of Astronomy, University of Texas, Austin, Texas 78712, USA}
\date{April 7, 2006}

\begin{abstract}
  We show that the energy density spectrum of the primordial gravitational waves 
  has characteristic features due to the successive changes in the relativistic 
  degrees of freedom during the radiation era. These changes make the evolution of
  radiation energy density deviate from the conventional adiabatic evolution, 
  $\rho_r\propto a^{-4}$, and thus cause the expansion rate of the universe to 
  change suddenly at each transition which, in turn, modifies the
 spectrum of primordial gravitational waves. We take into account all the particles in 
  the Standard Model of elementary particles. In addition, free-streaming of 
  neutrinos damps the amplitude of gravitational waves, leaving characteristic 
  features in the energy density spectrum. Our calculations are solely based on 
  the standard model of cosmology and particle physics, and therefore these features must exist. Our calculations significantly 
  improve the previous ones which ignored these effects and predicted 
  a smooth, featureless spectrum. 
\end{abstract}
\maketitle

\section{Introduction}%

Detection of the stochastic background of primordial gravitational waves has 
profound implications for the physics of  the early universe and the high energy 
physics that is not accessible by particle accelerators \cite{Grishchuk75, Starobinsky79, RebakovSV82,Fabri&Pollock83, Abbott&Wise84, Starobinsky85, Ford87, Allen88, Sahni90, TurnerWL, White, Turner, Kamionkowski}. 
The basic reason why relic gravitational waves carry information about the very early universe is that particles which decoupled from the primordial plasma at a certain time, $t\sim t_{\rm{dec}}$, when the universe had a temperature of $T\sim T_{\rm{dec}}$, memorize the physical state of the universe at and below $T_{\rm{dec}}$. Since gravitons decoupled below the Plank energy scale, the relic gravitons memorize all the expansion history of the universe after they decoupled and thus would probe deeper into the very early universe. Gravitational waves act therefore as the \textit{time machine} that allows us to see through the entire history of the universe.  Another example of relic species is the cosmic microwave background (CMB) photons, which decoupled from matter at $T\sim 0.3$ eV and can trace the physical state of the universe back to 0.3 eV.  On the other hand, the primordial gravitational waves carry information on the state of the much earlier universe than the CMB photons do. 
 
The purpose of this paper is to study the evolution of primordial gravitational waves through changes in the physical conditions in the universe within the Standard Model of elementary particles and beyond. For instance, the quark gluon plasma (QGP) phase to hadron gas phase transition causes a sharp feature in the gravitational wave spectrum. The change of the number of relativistic degrees of freedom affects the Hubble rate by reducing the growth rate of the Hubble radius during the transition. 
Thus, the rate at which modes re-enter the horizon is changed during the transition and a step in the spectrum appears at frequencies on the order of the Hubble rate at the transition. For the QGP phase transition this frequency is $\sim 10^{-7}$ Hz today and the correction is about 30\% \cite{Schwarz}. 
Other large drops in the number of relativistic degrees of freedom occur at electron positron annihilation and possibly at the supersymmetry (SUSY) breaking. Since the gravitational wave spectrum is sensitive to the number of relativistic degrees of freedom, one can search for evidence of supersymmetry in the very beginning of the universe by looking at the relevant frequency region ($\sim 10^{-3}$ Hz).  For energy scales lower than neutrino decoupling ($\sim 2$ MeV \cite{DicusKGSTT82}) we shall also account for the damping effect from neutrino free-streaming.

The primordial gravitational wave spectrum will also provide us with information about inflation. The energy scale of inflation is directly related to the amplitude of the spectrum. The modes which re-entered the horizon during the radiation dominated epoch show a nearly scale invariant spectrum if we do not consider the change of the effective number of degrees of freedom. Typically the amplitude of the spectrum is of order $10^{-15}$ for $10^{16}$ GeV inflation energy scale in such a frequency region. Inflation ends when the inflaton decays into radiation and reheats the universe \cite{Guth81, Linde82, Albrecht&Steinhardt82}. 
The energy scale of reheating could be seen from the highest frequency end ($\sim k_{\rm{rh}}$) of a nearly scale invariant energy density spectrum of the primordial gravitational waves. The lowest frequency mode observable today corresponds to the horizon size today, and the interval between the lowest frequency and $k_{\rm{rh}}$ would give the number of $e$-holdings, which tells us the duration of inflation between the end of inflation and the time at which fluctuations having the wavelength of the current horizon size left the horizon during inflation.
The slope of the spectrum provides the power-law index of the tensor perturbation, $n_T$ \cite{TurnerWL, Turner}.  
$n_T= 0$ corresponds to a scale invariant power spectrum from de Sitter inflation. In a large class of inflationary models $|n_T|$ is not zero but much smaller than unity, and its determination constrains the inflationary models. As the effect of $n_T$ has been investigated by many authors, e.g. \cite{TurnerWL, White, Turner, SmithKC05, SmithPC06}, and is easy to include, we shall assume de Sitter inflation ($n_T=0$) throughout this paper. Our result is general and easily applicable to any kind of models which produce primordial tensor perturbations. (e.g. Ekpyrotic models \cite{KhouryOST01}).

The primordial gravitational waves not only test and probe the physics of inflation and reheating, but also can provide the tomography of the standard model of particle physics and models beyond. 
The study of its spectrum enables us to probe the very early universe in a truly transparent way.
The goal of this paper is to show how the constituents in the early
and very early universe would
affect the primordial gravitational wave spectrum, which is observable in
principle and may be observable in the future by the next generation
observational projects, such as the Big Bang Observer (BBO) proposed to NASA \cite{BBO}
and the DECIGO proposed in Japan \cite{DECIGO}.
We present a new, rigorous computation of the primordial gravitational wave spectrum from de Sitter inflation with the Standard Model of particle physics. It is easy to extend our results to non-de Sitter (e.g., slow-roll) inflation models.

The outline of this paper is as follows. In Sec. \ref{basics}, basics about the primordial gravitational waves from inflation are reviewed.
In Sec. \ref{gstar}, a crucial quantity during radiation domination, the
effective relativistic degrees of freedom, $g_*$, is introduced and
related to the primordial gravitational wave spectrum in heuristic and
intuitive manners to illustrate the underlying physics. In
Sec. \ref{prediction}, we give an improved calculation of the primordial
gravitational wave spectrum in the Standard Model, employing de Sitter
inflation. Our final results are summarized in Figs.~\ref{fig4} and
\ref{fig5}.  
In Appendix~\ref{sbessel} we give useful formulae for the Bessel type functions.
In Appendix~\ref{solution} we give analytical solutions of gravitational waves in some limiting cases.
We define energy density of gravitational waves in Appendix~\ref{omega}.
The effect of neutrino free-streaming on the spectrum is formulated and explained
in Appendix \ref{neutrinodamp}. The numerical solution to the
integro-differential equation is also presented. 
In Appendix~\ref{oscillation} we give more detailed analytical accounts of
numerical solutions of gravitational waves when the effective number of relativistic species changes.
Units are chosen as $c=\hbar=k_B=1$ and $\sqrt{8\pi G}$ is retained. Indices $\lambda,\,\mu,\,\nu,\ldots$ run from 0 to 3, and $i,\,j,\,k,\ldots$ run from 1 to 3. Over-dots are used for derivatives with respect to time throughout the paper. Primes are mainly used for derivatives with respect to conformal time, but sometimes with respect to arguments we are focusing on. Barred quantities are unperturbed parts of variables.

\section{Wave equation, power spectrum, and energy density}\label{basics}%

In this section we define the power spectrum, $\Delta^2_h(k)$, 
and relative spectral energy density, $\Omega_h(k)$, of the
gravitational wave background. We do this because some authors use
different conventions in the literature.
For tensor perturbations on an isotropic, uniform and flat background spacetime, the metric is given by
\begin{eqnarray}
\label{2.1}
ds^2&=&a^2(\tau)[-d\tau^2+(\delta_{ij}+h_{ij})dx^i dx^j], \\
\label{2.2}
g_{\mu\nu}&=&a^2(\tau)(\eta_{\mu\nu}+h_{\mu\nu}), \\
\textrm{where} \nonumber\\
\eta_{\mu\nu}&=&\mathrm{diag}(-1,1,1,1), \quad h_{00}=h_{0i}=0,\quad |h_{ij}|\ll 1.
\end{eqnarray}
Here and after we shall work in the transverse traceless (TT) gauge, which leaves only the tensor modes in perturbations, i.e. $h_{ij,j}=0$ and $h^{i}{}_{i}=0$. In the linear perturbation theory the TT metric fluctuations are gauge invariant \footnote{In classic references \cite{Bardeen, KodamaSasaki}, $h_{ij}=2H_{Tij}$ and $\Pi_{ij}=\bar{p}\pi_{Tij}$ for tensor perturbations, which are automatically gauge-invariant.}. We shall denote the two independent polarization states of the perturbation as $\lambda=+, \times$ and sometimes suppress them when causing no confusion. We decompose $h_{ij}$ into plane waves with the comoving wave number, $|\mathbf{k}|\equiv k$, as
\begin{eqnarray}
\label{2.3}
h_{ij}(\tau,\mathbf{x}) = \sum_{\lambda}\int\!\frac{d^3k}{(2\pi)^3} h_{\lambda}(\tau;\mathbf{k}) e^{i\mathbf{k}\cdot\mathbf{x}}\epsilon_{ij}^{\lambda},
\end{eqnarray}
where $\epsilon_{ij}^{\lambda}$ is the polarization tensor and $\lambda=+, \times$. The equation for the wave amplitude, $h_{\lambda}(\tau;\mathbf{k})\equiv h_{\lambda,\mathbf{k}}$, is obtained by requiring the perturbed metric [Eq.~(\ref{2.2})] to satisfy the Einstein equation to $\mathcal{O}(h)$. One finds that $\delta G_{ij}=8\pi G\delta T_{ij}$ in the linear order \cite{G&C} 
yields
\begin{eqnarray}
\label{2.4}
-\frac{1}{2}h_{ij;\nu}{}^{;\nu}= 8\pi G\Pi_{ij},
\end{eqnarray}
where $\Pi_{ij}(t,\mathbf{x})$ is the anisotropic part of the stress tensor, defined by writing the spatial part of the perturbed energy-momentum tensor as 
\begin{eqnarray}
\label{2.5}
T_{ij}=pg_{ij} + a^2\Pi_{ij},
\end{eqnarray}
where $p$ is pressure. For a perfect fluid
$\Pi_{ij}=0$, but this would not be true in general. In the cosmological
context, the amplitude of gravitational waves is 
affected by anisotropic stress when
neutrinos are freely streaming (less than $\sim 10^{10}$K) \cite{Vishniac82, Bond&Szalay83, Kasai&Tomita85, Rebhan&Schwarz94, Weinberg04, Pritchard,Dicus&Repko05, Bashinsky05}. 
As we only deal with tensor perturbations, $h_{ij}$, we may treat each component as a scalar quantity under general coordinate transformation, which means e.g. $h_{ij;\mu}=h_{ij,\mu}$. The left-hand side of Eq. (\ref{2.4}) becomes
\begin{eqnarray}
\label{2.6}
h_{ij;\nu}{}^{;\nu} &=& \bar{g}^{\mu\nu}(h_{ij,\mu\nu}-\Gamma^{\alpha}_{\mu\nu}h_{ij,\alpha}), \nonumber\\ 
&=& -\ddot{h}_{ij}+\bigg(\frac{\nabla^2}{a^2}\bigg)h_{ij}-\bigg(\frac{3\dot{a}}{a}\bigg)\dot{h}_{ij},
\end{eqnarray}
where $\Gamma^0_{0\nu}=\Gamma^0_{\mu 0}=0,\;\Gamma^0_{ij}=\delta_{ij}\dot{a}a$, and $\bar{g}^{ij}=a^{-2}\delta_{ij}$ have been used. Commas denote partial derivatives, while semicolons denote covariant derivatives in Eqs. (\ref{2.4}) and (\ref{2.6}).
Transforming this equation into Fourier space, one obtains
\begin{eqnarray}
\label{2.7}
\ddot{h}_{\lambda,\mathbf{k}}+\bigg(\frac{3\dot{a}}{a}\bigg)\dot{h}_{\lambda,\mathbf{k}}+\bigg(\frac{k^2}{a^2}\bigg)h_{\lambda,\mathbf{k}}=16\pi G\Pi_{\lambda,\mathbf{k}},
\end{eqnarray}
where the Laplacian $\nabla^2$ in the second term of (\ref{2.4}) has
been replaced by $-k^2$ in the third term of (\ref{2.7}). The second
term represents the effect of the expansion of the universe. Using
conformal time derivatives ( $'\equiv\frac{\partial}{\partial\tau}$), we
may obtain
\begin{eqnarray}
\label{2.8}
h''_{\lambda,\mathbf{k}}+\bigg(\frac{2a'}{a}\bigg)h'_{\lambda,\mathbf{k}}+k^2
h_{\lambda,\mathbf{k}}=16\pi Ga^2\Pi_{\lambda,\mathbf{k}}.
\end{eqnarray}
This is just the massless Klein-Gordon equation for a plane wave in an
expanding space with a source term. Thus, each polarization state of the wave behaves as a massless, minimally coupled, real scalar field, with a normalization factor of $\sqrt{16\pi G}$ relating the two. 

Next, let us consider the time evolution of the spectrum. 
After the fluctuations left the horizon, $k\ll aH$,
equation~(\ref{2.8}) would become
\begin{equation}
 \frac{h''_{\lambda,\mathbf{k}}}{h'_{\lambda,\mathbf{k}}}\approx
  -\frac{2a'}{a}, 
\end{equation}
whose solution is 
\begin{equation}
 h_{\lambda,\mathbf{k}}(\tau) = A + B\int^\tau \frac{d\tau'}{a^2(\tau')},
\end{equation}
where $A$ and $B$ are integration  constants. Ignoring the second term
that is a decaying mode, one finds that $h_{\lambda,\mathbf{k}}$
remains constant outside the horizon. 
Note that we have ignored the effect of anisotropic stress outside the
horizon, as this term is usually given by causal mechanism which must
vanish outside the horizon.
Therefore, one may write a general solution of $h_{\lambda,\mathbf{k}}$
at any time as 
\begin{eqnarray}
\label{2.0.2a}
h_{\lambda,\mathbf{k}}(\tau)&\equiv&h^{prim}_{\lambda,\mathbf{k}}
\mathcal{T}(\tau,k),
\end{eqnarray}
where $h^{prim}_{\lambda,\mathbf{k}}$ is the primordial gravitational
wave mode that left the horizon during inflation. 
The transfer function, $\mathcal{T}(\tau,k)$, then describes the
sub-horizon evolution of gravitational wave modes after the modes
entered the horizon. The transfer function is normalized such 
that $\mathcal{T}(\tau,k)\rightarrow 1$ as $k\rightarrow 0$.
The power spectrum of gravitational waves, $\Delta^2_h(k)$, 
may be defined as
\begin{equation}
 \langle h_{ij}(\tau,\mathbf{x}) h^{ij}(\tau,\mathbf{x})\rangle
= \int \frac{dk}{k}\Delta^2_h(\tau,k),
\end{equation}
which implies
\begin{equation}
 \Delta_h^2(\tau,k) = \frac{2k^3}{2\pi^2}
\sum_\lambda\langle|h_{\lambda,\mathbf{k}}(\tau)|^2\rangle.
\end{equation}
Using equation~(\ref{2.0.2a}), one may write the 
time evolution of the power spectrum as
\begin{eqnarray}
\label{2.0.3}
\Delta_h^2(\tau,k)\equiv
\Delta_{h,prim}^2\left[\mathcal{T}(\tau,k)\right]^2,
\end{eqnarray}
where
\begin{equation}
\Delta_{h,prim}^2 =
 \frac{2k^3}{2\pi^2}\sum_\lambda
\langle|h_{\lambda,{\mathbf k}}^{prim}|^2\rangle
=\frac{16}{\pi}\left(\frac{H_{\mathrm{inf}}}{m_{Pl}}\right)^2.
\end{equation}
We have used the prediction for the amplitude of gravitational waves 
from de-Sitter inflation at the last equality, and $H_{\rm inf}$ is the
Hubble constant during inflation. One may easily extend this result to
slow-roll inflation models.

The energy density of gravitational waves is given by the 0-0 component
of stress-energy tensor of gravitational waves:
\begin{equation}
 \rho_h(\tau) = \frac{\langle h'_{ij}(\tau,\mathbf{x})h^{ij}{}'(\tau,\mathbf{x})\rangle}{32\pi Ga^2(\tau)}.
\end{equation}
The relative spectral energy density, $\Omega_h(\tau,k)$, 
is then given by the Fourier
transform of energy density, $\tilde{\rho}_h(\tau)\equiv \frac{d\rho_h}{d\ln{k}}$, 
divided by the critical density of the universe, $\rho_{\rm cr}(\tau)$
(see Appendix~\ref{omega} for full derivation):
\begin{eqnarray}
\Omega_h(\tau,k)&\equiv&\frac{\tilde{\rho}_h(\tau,k)}{\rho_{\rm{cr}}(\tau)}=\frac{\Delta_{h,prim}^2}{12a^2(\tau)H^2(\tau)}
\left[\mathcal{T}'(\tau,k)\right]^2.
\label{eq:omegaexact}
\end{eqnarray}
Note that $\Omega_h(k)$ is often defined as 
$\Omega_h(\tau,k) =
\frac{\Delta_{h,prim}^2}{12a^2(\tau)H^2(\tau)}k^2\left[\mathcal{T}(\tau,k)\right]^2=\frac{k^2}{12a^2(\tau)H^2(\tau)}\Delta_{h}^2(\tau,k)$
in the literature~\cite{TurnerWL, SmithKC05, 300years}. 
This definition is not compatible with the 0-0 component of stress
energy tensor; however, it is a good approximation when the modes are deep
inside the horizon, $k\gg aH$. Let us briefly explain a relation between these
two definitions. 
The transfer function is usually given by Bessel type functions,
$\mathcal{T}(x)=\frac{1}{x^n}[Aj_n(x)+By_n(x)]$. The conformal time
derivative of the transfer function is thus given by 
$\frac{d}{d\tau}\mathcal{T}(x)=
-\frac{k}{x^n}\left[Aj_{n+1}(x)+By_{n+1}(x)\right]$, 
where $x\equiv k\tau$. 
Therefore, in the limit that the modes are deep inside the horizon,
$k\gg aH$, one obtains 
$\Omega_h(k) = 
\frac{\Delta_{h,prim}^2}{12a^2(\tau)H^2(\tau)}
\left[\mathcal{T}'(\tau,k)\right]^2\approx
\frac{\Delta_{h,prim}^2}{12a^2(\tau)H^2(\tau)}k^2
\left[\mathcal{T}(\tau,k)\right]^2$,
which agrees with the definition of $\Omega_h(k)$ in \cite{TurnerWL, SmithKC05, 300years}. 
The difference between $\Omega_h$ and $k^2\Delta_h^2$ would affect the
prediction only at the largest scales, where both the overall amplitude and phases are different.
(The phases are shifted by $\pi/2$.)

\begin{figure}
  \begin{center}
    \resizebox{100mm}{!}{\includegraphics{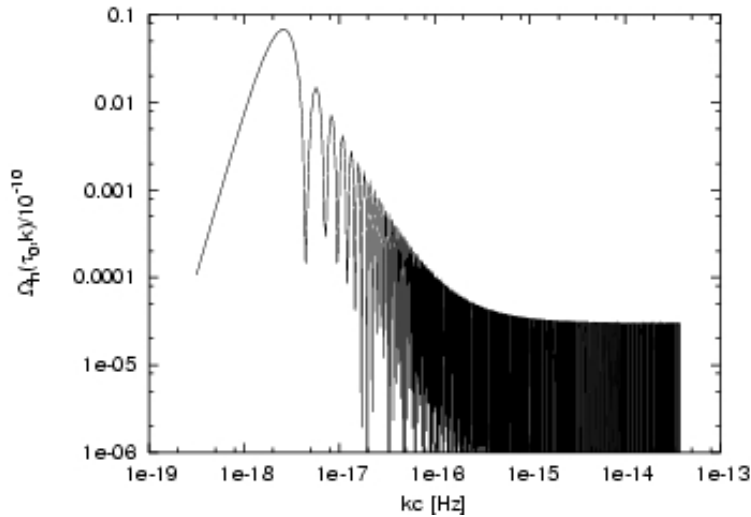}}
    \caption{The primordial gravitational wave spectrum at present,
   $\tau=\tau_0$, as a function of
   the comoving wavenumber, $k$ (or $kc$ in units of Hertz). The frequency of gravitational waves observed today is related to $k$ by $f_0=kc/2\pi$. The spectrum at large
   wavenumber is exactly scale-invariant as we have assumed de-Sitter 
   inflation. In this figure we have not taken into account the effects
   of the change in effective relativistic degrees of freedom or
   neutrino free-streaming.}
    \label{fig1}
  \end{center}
\end{figure}

Figure~\ref{fig1} shows a numerical calculation of 
equation~(\ref{eq:omegaexact}) for $\Omega_m=1-\Omega_r$, 
$\Omega_r h^2=4.15\times 10^{-5}$, and $h=0.7$. We ignored the
contribution from dark energy, which is only important 
at the lowest frequency regime that we are not interested in 
in this paper. 
One may understand the basic features in this numerical result as follows.
Energy density of gravitational waves evolves just like that of
radiation inside the horizon, $\tilde{\rho}_h(\tau,k)\propto a^{-4}$,
for $k\gg aH$. This implies that 
the relative spectral energy density,
$\Omega_h(\tau,k)$, inside the horizon remains independent of time during the
radiation era while it decreases as 
$\Omega_h(\tau,k)\propto a^{-1}$ during the
matter era. 
Therefore, the modes that entered the horizon during the matter era {\it
later} would decay {\it less}. As the low frequency modes represent the
modes that entered the horizon at late times, $\Omega_h(\tau,k)$ rises
toward lower frequencies. On the other hand, $\Omega_h(\tau,k)$  
at $k\gtrsim 10^{-15}$~Hz is independent of $k$. These are the modes
that entered the horizon during the radiation era for which
$\Omega_h(\tau,k)$ was independent of time. After the matter-radiation
equality all of these modes suffered the same amount of redshift, and
thus the shape of $\Omega_h(\tau,k)$ still remains scale-invariant at 
$k\gtrsim 10^{-15}$~Hz.

These qualitative arguments may be made more quantitative by using
the following analytical solutions of $\Omega_h(\tau,k)$ for three
different regimes (see Appendix~\ref{solution} for derivation):
\begin{eqnarray}
\label{2.1.10a}
\Omega_h(\tau<\tau_{\rm{eq}},k>k_{\rm{eq}})&=&\frac{\Delta_{h,prim}^2a^2}{12H^2_{\mathrm{eq}}a^4_{\mathrm{eq}}}k^2\left[j_1(k\tau)\right]^2,\\
\label{2.1.10aa}
\Omega_h(\tau>\tau_{\rm{eq}},k>k_{\rm{eq}})&=&\frac{\Delta_{h,prim}^2a}{12H_0^2a_0^3}k^2\frac{\tau^2_{eq}}{\tau^2}\left[A(k)j_2(k\tau)+B(k)y_2(k\tau)\right]^2,\\
\label{2.1.10b}
\Omega_h(\tau>\tau_{\rm eq},k<k_{\rm{eq}})&=&\frac{\Delta_{h,prim}^2a}{12H_0^2a_0^3}k^2\left[\frac{3j_2(k\tau)}{k\tau}\right]^2,
\end{eqnarray}
where $\tau_{\rm eq}$ is the conformal time at the matter-radiation
equality, $k_{\rm eq}$ is the comoving wavenumber of the modes that
entered the horizon at equality, and $j_0'(x)=-j_1(x)$ and
$\left[\frac{j_1(x)}{x}\right]'=-\frac{j_2(x)}{x}$ have been used to
compute $\mathcal{T}'(\tau,k)$. 
(Spherical Bessel functions are given in Appendix \ref{sbessel}.)
The first solution [Eq.~(\ref{2.1.10a})] describes
$\Omega_h(\tau,k)$ during radiation era for the modes that entered the
horizon before $\tau_{\rm eq}$. This solution is of course not relevant
to what we observe today. (We do not live in the radiation era.)
The second [Eq.~(\ref{2.1.10aa})] 
and third [Eq.~(\ref{2.1.10b})] solutions describe $\Omega_h(\tau,k)$
during matter era for the modes that entered the horizon before and
after $\tau_{\rm eq}$, respectively. 
The $k$-dependent 
coefficients $A(k)$ and $B(k)$ are given in equation~(\ref{2.0.2d}) and
(\ref{2.0.2e}), respectively. 
While the expression is slightly complicated, one can find that 
the second solution is independent of $k$ when the
oscillatory part is averaged out, which explains a scale-invariant 
spectrum at high frequencies, $k>k_{\rm eq}\sim 10^{-15}$~Hz. 
On the other hand, the third solution gives 
$\Omega_h(\tau,k)\propto k^{-2}$, which explains the low frequency
spectrum.

Figure~\ref{fig1} (and its extension to slow-roll inflation
which yields a small tilt in the overall shape of the spectrum) has been
widely referred to as the prediction from the standard model of cosmology.
However, as we shall show in the subsequent sections, 
the standard model of cosmology actually yields
much richer gravitational wave spectrum with more characteristic
features in it.

\section{The effective relativistic degrees of freedom: $g_*$}\label{gstar}%

It is often taken for granted that energy density of the universe
evolves as $\rho\propto a^{-4}$ during the radiation era. This is
exactly what caused a scale invariant spectrum of $\Omega_h(k)$ at
$k>k_{\rm eq}$. However,  $\rho\propto a^{-4}$ does not always hold even
during the radiation era, as some particles would become non-relativistic
before the others and stop contributing to the radiation energy density.

During the radiation era many kinds of particles interacted
with photons frequently so that they were in thermal equilibrium.
In an adiabatic system, the entropy per unit comoving volume must be 
conserved \cite{EU};
\begin{eqnarray}
\label{3.0.1}
S(T)&=&s(T)a^3(T)=constant,\\
\textrm{where}\nonumber\\
s(T)&=&\frac{2\pi^2}{45}g_{*s}(T)T^3.\nonumber
\end{eqnarray}
The entropy density, $s(T)$, is given by the energy density and
pressure; $s=(\rho+p)/T$. The energy density and pressure in such a
plasma-dominant universe are given by 
\begin{eqnarray}
\label{3.0.2a}
\rho(T)&=&\frac{\pi^2}{30}g_*(T)T^4,\\
\label{3.0.2b}
p(T)&=&\frac13\rho(T),
\end{eqnarray}
respectively, where we have defined the ``effective number of
relativistic degrees of freedom'', $g_*$ and $g_{*s}$, 
following \cite{EU}. 
These quantities, $g_*$ and $g_{*s}$, count the (effective) number of relativistic species
contributing to the radiation energy density and entropy, respectively.
One may call either (or both) of the two the effective number of 
relativistic degrees of freedom. 
Equation~(\ref{3.0.1}) and (\ref{3.0.2a}) immediately imply
that energy density of the universe during the radiation era should evolve as
\begin{equation}
 \rho \propto g_*g_{*s}^{-4/3}a^{-4}.
\end{equation}
Therefore, unless $g_*$ and $g_{*s}$ are independent of time, 
the evolution of $\rho$ would deviate from $\rho\propto a^{-4}$.
In other words, the evolution of $\rho$ during the radiation era is
sensitive to how many relativistic species the universe had at a given
epoch. 
As the wave equation of gravitational waves contains
$(a'/a)h'_{\lambda,k}$, the solution of $h_{\lambda,k}$ would be affected
by $g_*$ and $g_{*s}$ via the Friedman equation:
\begin{eqnarray}
\label{3.0.4}
\frac{a'(\tau)}{a^2}=H_0\sqrt{\left(\frac{g_*}{g_{*0}}\right)\left(\frac{g_{*s}}{g_{*s0}}\right)^{-4/3}\Omega_r\left(\frac{a}{a_0}\right)^{-4}+\Omega_m\left(\frac{a}{a_0}\right)^{-3}}.
\end{eqnarray}

\begin{table}
\caption{Particles in the Standard Model and their mass and helicity states}
\begin{center}
\label{table_g_star}
\begin{tabular}{|c|c|c|}
  \hline
  particle          & rest mass [MeV] & the number of helicity states: $g_i$ \\
  \hline
  $\gamma$          & 0               & 2 \\
  $\nu$,$\bar{\nu}$ & 0               & 6 \\
  $e^+$,$e^-$       & 0.51            & 4 \\
  $\mu^+$,$\mu^-$   & 106             & 4 \\
  $\pi^+$,$\pi^-$   & 135             & 2 \\
  $\pi^0$           & 140             & 1 \\
  $gluons$          & 0               & 16\\
  $u$,$\bar{u}$     & 5               & 12\\
  $d$,$\bar{d}$     & 9               & 12\\
  $s$,$\bar{s}$     & 115             & 12\\
  $c$,$\bar{c}$     & 1.3$\times 10^3$& 12\\
  $\tau^+$,$\tau^-$ & 1.8$\times 10^3$& 4 \\
  $b$,$\bar{b}$     & 4.4$\times 10^3$& 12\\
  $W^+$,$W^-$       & 80$\times 10^3$ & 6 \\
  $Z$               & 91$\times 10^3$ & 3 \\
  $H$               & 114$\times 10^3$& 1 \\
  $t$,$\bar{t}$     & 174$\times 10^3$& 12\\
  SUSY particles    &$\sim1\times 10^6$&$\sim110$\\ 
  \hline
\end{tabular}
\end{center}
\end{table}

\begin{figure}
  \begin{center}
    \resizebox{100mm}{!}{\includegraphics{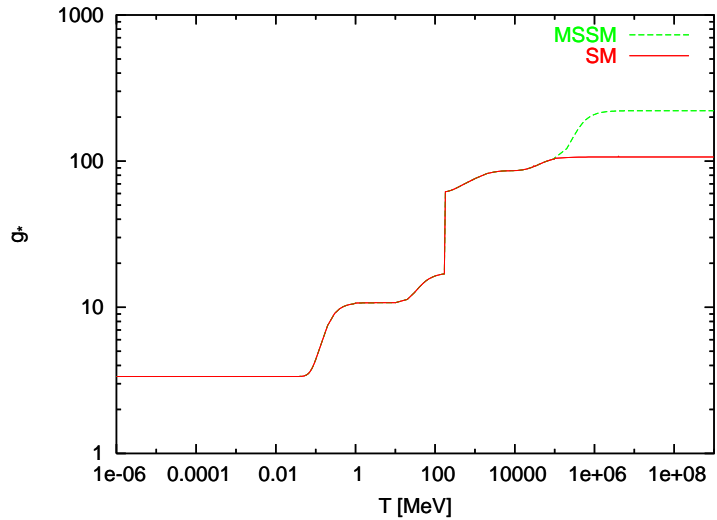}}
    \caption{Evolution of the effective number of relativistic degrees
   of freedom contributing to energy density, $g_*$, as a function of
   temperature. The solid and dashed lines represent $g_*$ in the
   Standard Model and in the minimal extension of Standard Model,
   respectively. At the energy scales above $\sim 1$ TeV, $g_*^{\rm{SM}}=106.75$
   and $g_*^{\rm{MSSM}}\simeq 220$. At the energy scales below $\sim 0.1$
   MeV, $g_*=3.3626$ and $g_{*s}=3.9091$; $g_*=g_{*s}$ otherwise.}
    \label{g_star}
  \end{center}
\end{figure}
\begin{figure}
  \begin{center}
    \resizebox{100mm}{!}{\includegraphics{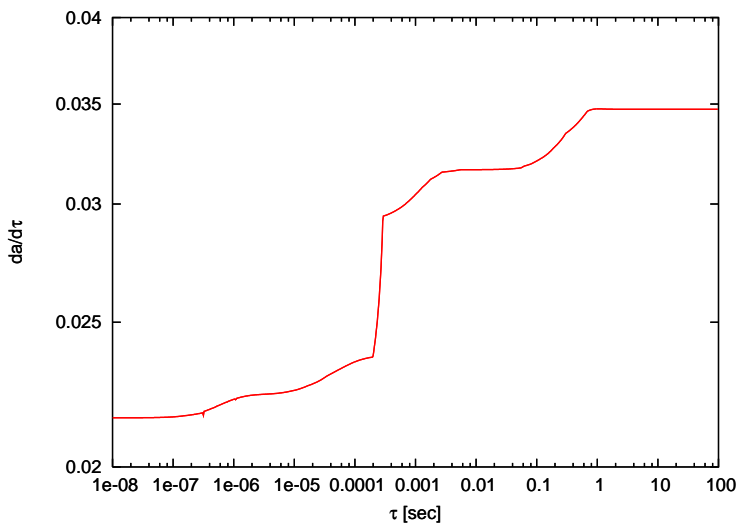}}
    \caption{Evolution of $a'$ as a function of the conformal time. If 
$g_*$ and $g_{*s}$ were constant, $\rho\propto a^{-4}$ and $a'$ would
   also be constant.}
    \label{a0}
  \end{center}
\end{figure}

Although the interaction rate among particles and antiparticles is
assumed to be fast enough (compared with the expansion rate) to keep them in thermal equilibrium, the
interaction is assumed to be weak enough for them to be treated as 
ideal gases. In the case of an ideal
gas at temperature $T$, each particle species of a given mass, 
$m_i=x_iT$, would contribute to $g_*$ and $g_{*s}$ the amount given by
\begin{eqnarray}
\label{3.0.3a}
g_{*,i}(T)&=&g_i\frac{15}{\pi^4}\int^{\infty}_{x_i}\frac{(u^2-x_i^2)^{1/2}}{e^u\pm1}u^2du,\\
\label{3.0.3b}
g_{*s,i}(T)&=&g_i\frac{15}{\pi^4}\int^{\infty}_{x_i}\frac{(u^2-x_i^2)^{1/2}}{e^u\pm1}(u^2-\frac{x_i^2}{4})du,
\end{eqnarray}
where the sign is $+$ for bosons and $-$ for fermions. 

Here, $g_i$ is the number of helicity states  of the particle and 
antiparticle. Note that an integral variable is defined as $u\equiv
E/T$, where $E=\sqrt{|\mathbf{p}|^2+m^2}$. We assume that the chemical potential,
$\mu_i$, is negligible. 
One might also define a similar quantity for the number density,
\begin{equation}
 n(T) = \frac{\zeta(3)}{\pi^2}g_{*n}T^3,
\end{equation}
where  $\zeta(3)\simeq 1.20206$ is the Riemann zeta function of 3.
Each species would contribute to $g_{*n}$ by 
\begin{equation}
 g_{*n,i}(T)=g_i\frac{1}{2\zeta (3)}\int^{\infty}_{x_i}\frac{(u^2-x_i^2)^{1/2}}{e^u\pm1}udu.
\end{equation}
The effective number relativistic degrees of freedom is then given by
the temperature-weighted sum of all particle contributions:
\begin{eqnarray}
\label{3.0.3c}
g_*(T)=\sum_{i}
g_{*,i}(T)\left(\frac{T_i}{T}\right)^4,\qquad g_{*s}(T)=\sum_i g_{*s,i}(T)\left(\frac{T_i}{T}\right)^3,\qquad 
g_{*n}(T)=\sum_i g_{*n,i}(T)\left(\frac{T_i}{T}\right)^3,
\end{eqnarray}
where we have taken into account the possibility that each species $i$
may have a thermal distribution with a different temperature 
from that of photons. The most famous example is neutrinos: 
$T_{\nu}=\left(\frac{4}{11}\right)^{1/3}T_{\gamma}$.
Neutrinos are cooler than photons at temperatures below MeV scale 
due to photon heating from electron-positron annihilation.

Fig. \ref{g_star} shows the evolution of $g_*$ as a function of 
temperature. 
We have included all the particles in the Standard Model of elementary
particles, as listed in table \ref{table_g_star}.
(Note that we assume that the mass of Higgs bosons is 114~GeV, which is
the current lower bound from experiments.) 
We neglected hadrons whose mass is heavier than pions. 
In addition to the particles in the Standard Model, one may also include 
particles in supersymmetric models. 
Superpartners in the minimal extension of supersymmetric Standard Model
(MSSM) would carry almost the same $g_*$ as that carried by particles 
within the Standard Model.
Fig. \ref{a0} shows the evolution of $a'$. If $g_*$ and $g_{*s}$ were
constant, $a'$ would also be constant during the radiation era; however, 
the evolution of $a'$ reveals a series of jumps due to the change in 
$g_*$ and $g_{*s}$.

Interactions between particles would change the ideal gas result
obtained above, and one cannot use equation~(\ref{3.0.3a}), 
(\ref{3.0.3b}) and (\ref{3.0.3c}) to calculate $g_*$ or $g_{*s}$.
Instead, one needs to extract $g_*$ and $g_{*s}$ directly from
energy density and entropy which would be calculated using detailed 
numerical simulations of particle interactions.
For example, above the critical temperature of Quark Gluon Plasma (QGP)
phase transition, most of $g_*$ is carried by color degrees of freedom. 
The dominant correction therefore comes from the colored sector of the
Standard Model, whereas corrections from the weak charged sector are suppressed
by the masses of weak gauge bosons. Since physics of QCD correction is
still uncertain and beyond the scope of this paper, we shall ignore this
effect and  treat it as an ideal gas case. 
The effects of particle interactions on 
$g_*$ have been investigated by 
\cite{DicusKGSTT82,ManganoMPP02,Hindmarsh}.

\subsection{Heuristic argument based on background density}\label{gstar1}%

Before presenting the full numerical results, let us briefly describe
how $g_*$ and $g_{*s}$ would affect the shape of $\Omega_h(\tau_0,k)$.
In Sec. \ref{basics}, we discussed how the expansion of the universe
would affect $\Omega_{h}(\tau_0,k)$. While energy density of the
universe during the radiation era is affected by $g_*$ and $g_{*s}$ as 
$\rho_{\rm cr}\propto g_*g_{*s}^{-4/3}a^{-4}$, energy density of 
gravitational waves always evolves as 
$\tilde{\rho}_h(\tau,k)\propto a^{-4}$ inside
the horizon, $k\gg aH$, regardless of $g_*$ or $g_{*s}$.
(Gravitons are not in thermal equilibrium with other particles.)
This difference in the evolution of $\tilde{\rho}_h$ and $\rho_{\rm cr}$ 
significantly modifies a scale-invariant spectrum of
$\Omega_h(\tau_0,k)$ at $k>k_{\rm eq}$.

Let us consider a gravitational wave mode with $k$ 
which entered the horizon 
at a given time, $\tau_{\rm{hc}}<\tau_{\rm eq}$ 
and temperature, $T=T_{\rm{hc}}$,
 during the radiation era. After the mode entered the horizon the amplitude
 of this mode would be suppressed by the cosmological redshift. 
The relative spectral density at present would then be given by
\begin{eqnarray}
\label{3.1.1}
\Omega_{h}(\tau_0,k>k_{\rm{eq}})
&=&\Omega_{h}(\tau_{\rm{hc}},k)\Omega_{r0}\left[\frac{g_{*s}(T_{\rm{hc}})}{g_{*s0}}\right]^{-4/3}
\left[\frac{g_{*}(T_{\rm{hc}})}{g_{*0}}\right],
\end{eqnarray}
where $\Omega_{r}$ denotes the relative energy density of radiation and the subscript ``0'' denotes the present-day value. 
This equation helps us understand how $g_*$ and $g_{*s}$ would affect
$\Omega_h(\tau_0,k)$. For a given wavenumber, $k$, there would be one
horizon-crossing epoch, $\tau_{\rm hc}$. 
The amount by which the relative spectral
energy density of that mode would be suppressed depends on
$g_*$ and $g_{*s}$ at $\tau_{\rm hc}$.
The mode that entered the horizon earlier should experience larger
suppression, as $g_*$ and $g_{*s}$ would be larger than those for the
mode that entered the horizon later. (The effective number of
relativistic degrees of freedom is larger at earlier times --- see
Figure~\ref{g_star}.) As $g_*$ and $g_{*s}$ are equal for 
$T\gtrsim 0.1$~MeV and nearly the same otherwise 
($g_*=3.3626$ and $g_{*s}=3.9091$ for $T\lesssim 0.1$~MeV), 
we expect that suppression factor is given by 
$(g_*/g_{*0})^{-1/3}$ to a good approximation.
The modes that entered the horizon during the matter era should not
be affected by $g_*$ or $g_{*s}$, as they do not change during the
matter era.

\subsection{More rigorous argument using analytical solutions}\label{gstar2}%

In this subsection we derive equation~(\ref{3.1.1}) using a more
rigorous approach. 
Let us go back to the wave equation [Eq.~(\ref{2.8})], and rewrite it
using a new field variable, $\mu_k\equiv a h_k$: 
\begin{eqnarray}
\label{3.1}
\mu''_{k}+\bigg(k^2-\frac{a''}{a}\bigg)\mu_{k}=16\pi Ga^3\Pi_{k}.
\end{eqnarray}
Note that we have suppressed the subscript for polarization, $\lambda$.
To find a solution for $\mu_k$, we must solve the Friedman equation as 
well:
\begin{eqnarray}
\label{3.2.2}
\left(\frac{a'}{a^2}\right)^2&=&\frac{8\pi
G}{3}\rho_r=H_0^2\frac{g_*}{g_{*0}}\left(\frac{g_{*s}}{g_{*s0}}\right)^{-4/3}\left(\frac{a}{a_0}\right)^{-4}\\
\label{3.2.3}\nonumber
\frac{a'}{a^2}&=&H_0\left(\frac{g_{*}}{g_{*0}}\right)^{1/2}\!\!\left(\frac{g_{*s}}{g_{*s0}}\right)^{-2/3}\!\!\left(\frac{a}{a_0}\right)^{-2}\\
\label{3.2.3a}\nonumber
\frac{a-a_0}{a_0}&=& a_0H_0 \int^{\tau}_{\tau_0}\!\!d\tau'\left(\frac{g_{*}}{g_{*0}}\right)^{1/2}\!\!\left(\frac{g_{*}}{g_{*0}}\right)^{-2/3},
\end{eqnarray}
where the subscript ``0'' denotes some reference epoch during the radiation
era. (While ``0'' means the present epoch in the other sections, we use
it to mean some epoch during the radiation era in this section only.)
To proceed further, we need to specify the evolution of $g_*$ and
$g_{*s}$. While we have numerical data for the evolution of these
quantities, we make an approximation here to make the problem 
analytically solvable.
Since $g_*(\tau)$ decreases monotonically as the universe expands, 
one may try a reasonable {\it ansatz}, $g_* \propto \tau^{-6n}$, to obtain
analytical solutions. 
We shall also assume $g_*= g_{*s}$ and $\Pi_{k}=0$ for simplicity
in this section. (At temperatures below 2 MeV, free-streaming of
neutrinos generates anisotropic stress, $\Pi_{k}\neq 0$.
Also, the temperature of neutrinos is different from that of photons
below electron-positron annihilation temperature, and thus 
$g_*\neq g_{*s}$ below $\sim 0.51$ MeV.)
This model gives
\begin{eqnarray}
\label{3.2.4}
\frac{a''}{a}&\simeq&\frac{g_*^{-7/6}\mid g_*'\mid}{6\int^{\tau}_{\tau_0}d\tau'g_*^{-1/6}} \simeq \frac{n}{\tau^2}\left(1+n\right),
\end{eqnarray}
where the primes denote derivatives with respect to $\tau$, and the term
$\frac{n}{\tau_0^2}\left(1+n\right)$ has been neglected in the last
line, assuming $\tau\ll \tau_0$.
This form of $a''/a$ allows us to find an analytical solution to 
equation~(\ref{3.1}):
\begin{eqnarray}
\label{3.2.5}
h_k(\tau)=\frac{\mu_k(\tau)}{a}=A(k)\frac{j_n(k\tau)}{(k\tau)^{n}}+B(k)\frac{y_n(k\tau)}{(k\tau)^{n}},
\end{eqnarray}
where $A(k)$ and $B(k)$ are the normalization constants that should be determined
by the appropriate boundary conditions. 
Note that $n=0$ and $n=1$ correspond to the solutions for the radiation era
and the matter era, respectively.

Let us consider a model of the radiation-dominated universe in which 
there was a brief period of time during which $g_*$ suddenly decreased as
a power-law in time, $g_*\propto \tau^{-6n}$. 
Outside of this period $g_*$ is a constant.
Suppose that $g_*$ changed between $\tau=\tau_2$ and $\tau_1>\tau_2$.
(The change in $g_*$ began at $\tau=\tau_2$ and completed at $\tau_1$.)
The modes that entered the horizon after $\tau_1$ do not know
anything about the change in $g_*$. The solution for such modes 
is therefore given by the usual solution during the radiation era,
\begin{eqnarray}
\label{3.2.6a}
h^{\rm{out}}_k(\tau)&=&h_k^{prim} j_0(k\tau).
\end{eqnarray}
How about the modes that entered the horizon before $\tau_2$? 
The solution for such modes is given by
\begin{eqnarray}
\label{3.2.6b}
h^{\rm{in}}_k(\tau)&=&h_k^{prim} j_0(k\tau) \qquad\qquad\qquad\qquad (\tau<\tau_2), \\
\label{3.2.6c}
h^{\rm{in}}_k(\tau)&=&\frac{h_k^{prim}}{(k\tau)^{n}} [A(k) j_{n}(k\tau)+B(k) y_{n}(k\tau)]
\qquad (\tau_2<\tau<\tau_1), \\
\label{3.2.6d}
h^{\rm{in}}_k(\tau)&=&h_k^{prim} \left(\frac{\tau_2}{\tau_1}\right)^{n}[C(k) j_0(k\tau)+D(k) y_0(k\tau)] \quad (\tau_1<\tau).
\end{eqnarray}
It is convenient to define $\tau_*\equiv(\tau_1+\tau_2)/2$ and 
$\Delta\tau\equiv\tau_1-\tau_2$ to characterize the time of 
transition and its duration, respectively.
Here, the superscript ``in'' denotes the modes that have already been
\textit{inside} the horizon at $\tau_*$, while ``out'' denotes the modes
that are still \textit{outside} the horizon at $\tau_*$. 
The coefficients, $A(k)$, $B(k)$, $C(k)$, and $D(k)$, are given 
by Eqs. (\ref{A.1.1}) -- (\ref{A.1.4}) in Appendix~\ref{oscillation}.
By taking a ratio of equation~(\ref{3.2.6d}) and (\ref{3.2.6a}), we can
find the amount of suppression in $h_k^{\rm{in}}(\tau>\tau_1)$ 
relative to $h_k^{\rm{out}}(\tau>\tau_1)$:
\begin{eqnarray}
\label{3.2.7}
\frac{h^{\rm{in}}_{k}(\tau>\tau_1)}{h^{\rm{out}}_{k}(\tau>\tau_1)} 
&=& \left(\frac{\tau_2}{\tau_1}\right)^{n}
[C(k)+D(k)y_0(k\tau)/j_0(k\tau)]
\approx \left(\frac{\tau_2}{\tau_1}\right)^{n}[C(k)+D(k)],
\end{eqnarray}
where we have ignored the oscillatory part of 
$y_0(k\tau)/j_0(k\tau)$. 
While $C(k)$ and $D(k)$ have fairly cumbersome expressions,
the sum of the two has a simple limit, $[C(k)+D(k)]^2\to1$, for 
$\Delta\tau \to 0$, regardless of the value of $n$ 
(see Appendix~\ref{oscillation}). 
The energy density in gravitational waves then reflects the effect from the change of $g_*$ as
\begin{eqnarray}
\label{3.2.8}
\frac{\Omega^{\rm{in}}_{h}(\tau>\tau_1,k)}{\Omega^{\rm{out}}_{h}(\tau>\tau_1,k)}\approx
\left[\frac{h^{\rm{in}}_{k}(\tau>\tau_1)}
{h^{\rm{out}}_{k}(\tau>\tau_1)}\right]^2
&\approx& 1-2n\frac{\Delta\tau}{\tau_1},
\end{eqnarray}
where we have used the sub-horizon limit for $\Omega_h(k)$ and 
$\Delta\tau\ll \tau_2<\tau_*<\tau_1$.
On the other hand, $g_* \propto \tau^{-6n}$ gives
\begin{eqnarray}
\label{3.2.9}
\frac{g^{\rm{in}}_{*}}{g^{\rm{out}}_{*}}\equiv
\frac{g_{*}(\tau_2)}{g_{*}(\tau_1)}
=\left(\frac{\tau_*-\Delta\tau/2}{\tau_*+\Delta\tau/2}\right)^{-6n}
\approx 1+6n\frac{\Delta\tau}{\tau_*}.
\end{eqnarray}
Hence, combining Eqs.(\ref{3.2.8}) and (\ref{3.2.9}), we finally obtain 
the desired result 
\begin{eqnarray}
\label{3.2.10}
\frac{\Omega^{\rm{in}}_{h}(k)}{\Omega^{\rm{out}}_{h}(k)}\approx\left(\frac{g^{\rm{in}}_{*}}{g^{\rm{out}}_{*}}\right)^{-1/3},
\end{eqnarray}
for $\Delta\tau \ll \tau_*$. 
This result agrees with equation~(\ref{3.1.1}), which was obtained
in the previous section (Sec.~\ref{gstar1}) using a more heuristic 
argument. (Note that we have assumed $g_*=g_{*s}$ in this section). 
In Eq.~(\ref{3.1.1}) there is an extra factor $\Omega_{r0}$, 
which represents the time evolution 
of $\Omega_h$ from matter-radiation equality to the present epoch.
We do not have this factor in equation~(\ref{3.2.10}), 
as both $\Omega_h^{\rm{in}}$ and
$\Omega_h^{\rm{out}}$ are evaluated during the radiation era.

\section{Prediction for Energy Density of Gravitational Waves from the Standard Model and beyond}\label{prediction}%

In Sec.~\ref{gstar} we have described how the evolution of the
effective number of relativistic degrees of freedom would affect the
shape of relative spectral energy density of primordial gravitational
waves at present, $\Omega_h(\tau_0,k)$. In this section we present the full
calculation of $\Omega_h(\tau_0,k)$, numerically integrating the wave
equation together with the numerical data of $g_*$ and $g_{*s}$ (see
Figure~\ref{g_star}). 

Before we do this, there is another effect that one must take into
account. While we have ignored anisotropic stress on the right hand side
of the wave equation~(\ref{2.8}) so far, free-streaming of relativistic neutrinos 
which have decoupled from thermal equilibrium at $T\lesssim 2$~MeV
significantly contributes to anisotropic stress, damping the amplitude of
primordial gravitational waves \cite{Weinberg04, Pritchard}.
Calculations given in Appendix~\ref{neutrinodamp} show that 
neutrino anisotropic stress damps $\Omega_h(\tau_0,k)$ by 35.5\%
in the frequency region between  $\simeq 10^{-16}$ and 
$\simeq 2\times 10^{-10}$~Hz.
The damping effect is much less significant below $10^{-16}$~Hz, as
this frequency region probes the universe that is dominated by matter.
One may understand this by looking at the right hand side of Eq. (\ref{4.1}). 
Anisotropic stress is proportional to the fraction of the total energy 
density in neutrinos, $f_{\nu}(\tau)$, which is very small when the
universe is matter dominated.

\begin{figure}
  \begin{center}
    \resizebox{110mm}{!}{\includegraphics{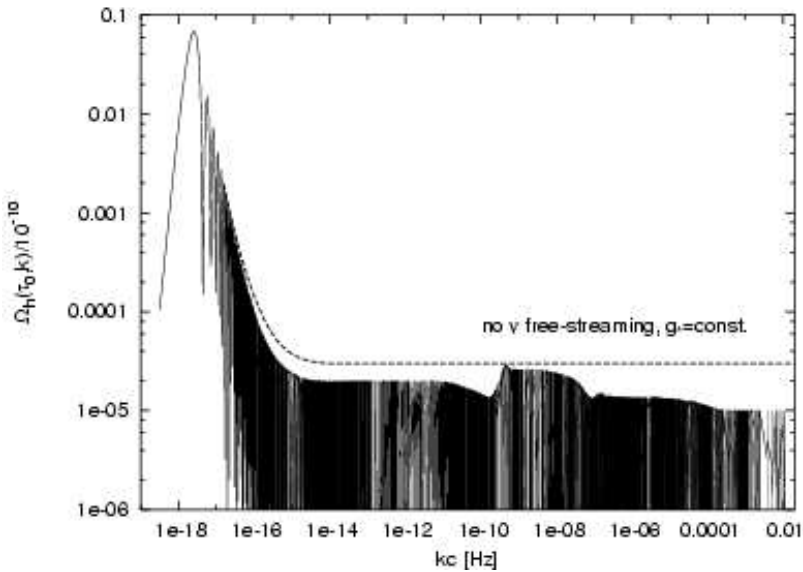}}
    \caption{The primordial gravitational wave spectrum at present,
   $\Omega_h(\tau_0,k)/10^{-10}$, as a function of
   the comoving wavenumber, $k$ (or $kc$ in units of Hertz). The frequency of gravitational waves observed today is related to $k$ by $f_0=kc/2\pi$. 
   We have assumed a scale-invariant primordial spectrum and
   $\Omega_m=1-\Omega_r$, $\Omega_r=4.15\times 10^{-5}h^{-2}$, $h=0.7$,
   and $E_{\rm{inf}}=10^{16}$ GeV. We have included the effects of
   the effective number of relativistic degrees of freedom and neutrino
   free-streaming. The dashed line shows the envelope of the previous calculations which ignored the change in the number of relativistic degrees of freedom and
   neutrino free-streaming (Fig.~\ref{fig1}).}
    \label{fig4}
  \end{center}
\end{figure}

\begin{figure}
  \begin{center}
    \resizebox{110mm}{!}{\includegraphics{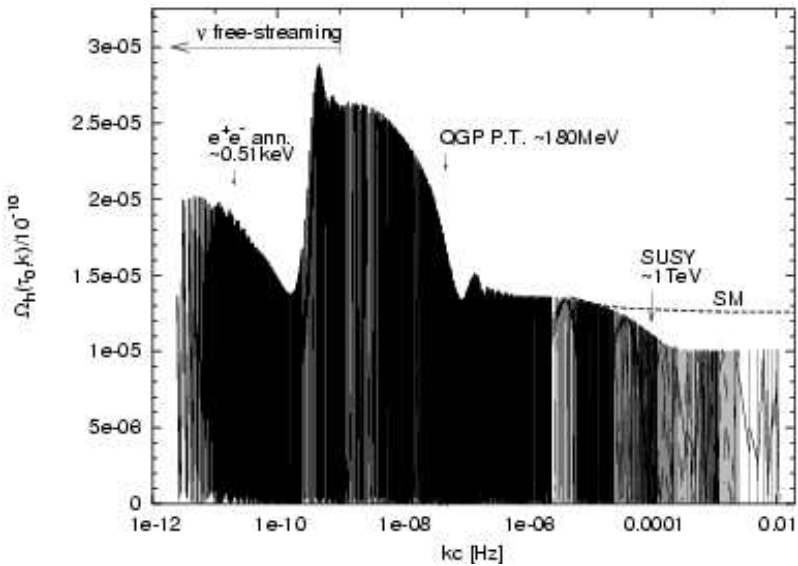}}
    \caption{A blow-up of Fig.~\ref{fig4}.
    Note that density of vertical lines shows density of sampling points
   at which we evaluate $\Omega_h(\tau_0,k)$. The dashed line shows the envelope of the spectrum in the Standard Model of elementary particles.}
    \label{fig5}
  \end{center}
\end{figure}

We show the results of full numerical integration in
Figure~\ref{fig4} and \ref{fig5}. The latter figure is
just a zoom-up of interesting features in the former one.  
We find that $\Omega_h(\tau_0,k)$ oscillates very rapidly as
$\sin^2{(k\tau+\varphi)}$, where $\varphi$ is a phase constant. The
cross term, $\sin{k\tau}\cos{k\tau}$, appeared as a beat in  
Fig.~\ref{fig1}, while they are too small to see in
Fig~\ref{fig4}. From observational point of view these oscillations will not be
detectable, as observations are only sensitive to the average power over
a few decades in frequency.

The damping effect due to neutrino free-streaming is evident below
$2\times 10^{-10}$, while 
one might also notice a minor wiggly feature at around $5\times 10^{-10}$~Hz. 
This feature is actually artificial. We implicitly assumed an
instantaneous decoupling of neutrinos from the thermal plasma at 
$T_{\nu\,\rm{dec}}=2$ MeV, which resulted in the \textit{surface} of
decoupling that is extremely thin. This gave rise to dips and peaks
corresponding to the waveform of gravitational waves at the decoupling 
time. (The envelope shape is somewhat similar to $-j_1(k)$ at 
around $5\times 10^{-10}$ Hz; more details are given in 
Appendix \ref{neutrinodamp}.) 
Physically speaking, however, the last scattering surface of neutrinos
is very thick, unlike for photons. (There is no ``recombination'' for
neutrinos.) Therefore, the oscillatory feature would be smeared out when
thickness of the decoupling surface is explicitly taken into account. To
do this, one would need to solve the Boltzmann equation for neutrinos
separately, including the effect of neutrino decoupling.

The effect of evolution of $g_*$ and $g_{*s}$ is also quite prominent. 
For example, big changes in $g_*$ would occur at the electron-positron
annihilation epoch, $\sim 0.51$~MeV ($\sim 2\times 10^{-11}$~Hz), as
well as at the QGP to hadron gas phase transition epoch, $\sim 180$~MeV ($\sim
10^{-7}$~Hz) within the Standard Model.  
The gravitational wave spectrum is suppressed by roughly 20\% and 30\% above
the electron-positron annihilation and QGP phase transition scale,
respectively. 
If supersymmetry existed above a certain energy scale, e.g., $\sim 1$ TeV
($\sim 1\times 10^{-4}$Hz), the spectrum would be suppressed by at least $\sim
20\%$ (for N=1 supersymmetry) above that frequency.
We also find additional features at the QGP phase transition scale,
$\sim 10^{-7}$~Hz, similar to the features at $\sim 5\times 10^{-10}$~Hz
caused by our assumption about instantaneous decoupling of neutrinos.
The feature at the QGP phase transition is nevertheless not artificial
--- as the QGP phase transition is expected to have happened in a short
time period, the instantaneous transition would be a good approximation,
unlike for neutrinos.

One may approximately relate the horizon crossing temperature of the universe to the
frequency of the gravitational waves \cite{Kamionkowski,Maggiore}. 
The horizon crossing mode, $k_{\rm{hc}}=a_{\rm{hc}}H_{\rm{hc}}$, is
related to the temperature at that time by
$H^2_{\rm{hc}}=\frac{8\pi^3G}{90}g_{*,\rm{hc}}T^4_{\rm{hc}}$. 
Then using entropy conservation,
$g_{*s,\rm{hc}}a^3_{\rm{hc}}T^3_{\rm{hc}}=g_{*s0}a^3_{0}T^3_{0}$, one
obtains the following conversion factor from the temperature of the
universe to the frequency of gravitational waves observed today:
\begin{eqnarray}
\label{5.1}
f_0= 1.65\times 2\pi \times 10^{-7}\left(\frac{T_{\rm{hc}}}{1\rm{GeV}}\right)
\left[\frac{g_{*s}(T_{\rm{hc}})}{100}\right]^{-1/3} 
\left[\frac{g_*(T_{\rm{hc}})}{100}\right]^{1/2} \textrm{Hz},
\end{eqnarray}
which was derived in \cite{Kamionkowski,Maggiore}. (If we take
$\epsilon \equiv \frac{1}{2\pi}$ in \cite{Maggiore}, their equation (156)
agrees with the one above.) 
Throughout this paper we have been using the comoving wavenumber, $k$ (or $kc$ in units of Hertz), which is related
to the conventional frequency by  $2\pi f_0=kc/a_0$, where $a_0$ is the
present-day scale factor and $c$ is the speed of light.
We use $k$ in this paper, rather than $f_0$, as $k$ is what enters into the wave equation
that we solve numerically.


\section{Discussion and Conclusion}\label{D&C}%

We have calculated the primordial gravitational wave spectrum, fully taking into account the evolution of the effective relativistic degrees of freedom and neutrino free-streaming, which were ignored in the previous calculations. 
The formalism and results given in this paper are based on solid physics and can be extended to primordial gravitational waves produced in any inflationary models and high energy particle physics models. 
As is seen in Figs.~\ref{fig4} and \ref{fig5}, the spectrum is no longer scale invariant, but has complex features in it. Whatever physics during inflation is, one must include the evolution of the effective relativistic degrees of freedom and neutrino free-streaming.

\cite{Schwarz} studied the gravitational wave spectrum at the QGP phase
transition assuming the first order instantaneous model as well as 
the second order cross-over model, and found 30\% suppression of the
energy density spectrum, which is consistent with our calculation. 
\cite{Seto&Yokoyama03} studied the effect of entropy
production from e.g., decay of massive particles in the early universe
on the energy density spectrum.
We have not included this effect in our calculations, as the late-time
entropy production is not predicted within the Standard Model.
\cite{Boyle&Steinhardt05} studied the effect of changes in  the
equation of state of the universe on the energy density spectrum.
While they included the effect of neutrino free-streaming, they did not
include the evolution of $g_*$. Instead, they explored general
possibilities that the equation of state might be modified by 
trace anomaly or interactions among particles. They also considered 
damping of gravitational waves due to anisotropic stress of some
hypothetical particles. Our calculations are different from theirs, as 
we took into account explicitly all the particles in the Standard Model
and the minimal extension of the Standard Model, but did not include any
exotic physics beyond that.

Let us mention a few points that would merit further studies. At the
energy scales where supersymmetry is unbroken (if it exists), say TeV
scales and above, the number of relativistic degrees of freedom, $g_*$, should be
at least doubled, and would cause suppression of the primordial
gravitational waves (Fig.~\ref{fig5} for $N=1$ supersymmetry).
If, for instance, $N=8$ is the number of internal supersymmetric
charges, $\sim 250$ copies of standard model particles would appear in this theory. 
This would suppress the spectrum by 85\% at the high frequency region (above $\sim 10^{-4}$ Hz) compare to the Standard Model, though the details depend on  models.  
Since we still do not have much idea about a \textit{true} supersymmetric model and its particle rest mass, the search for the primordial gravitational waves would help to constrain the effective number of relativistic degrees of freedom $g_*(T)$ above the TeV scales.

In an extremely high frequency region, $k_{\rm{rh}}$, the gravitational wave spectrum
should provide us with unique information about the reheating of the
universe after inflation. 
If the inflaton potential during reheating is monomial, $V(\phi)\propto
\phi^n$, the equation of state during reheating is given by
$p_{\phi}=\alpha(n)\rho_{\phi}$, where $\alpha(n)=\frac{n-2}{n+2}$. 
Since the equation of state determines the expansion law of that epoch,
one obtains the frequency dependence of the gravitational wave spectrum
as $\Omega_{h}\propto k^{(n-4)/(n-1)}$. 
In an extremely low frequency region (below $\sim 10^{-18}$~Hz), on the
other hand, 
dark energy dominates the universe and affects the
spectrum. Acceleration of the universe reduces 
the amplitude of gravitational waves that enter the horizon at this
epoch; however, we will not be able to observe modes as big as the size of the horizon today.

The signatures of the primordial gravitational waves may be detected
only by the CMB polarization in the low frequency region, $\lesssim
10^{-16}$~Hz. 
For the higher frequency region, however, direct detection of the
gravitational waves would be necessary, and it should allow us to search
for a particular cosmological event by arranging an appropriate
instrument, as the events during the radiation era are imprinted on the
spectrum of the primordial gravitational waves.

\appendix%

\section{Spherical Bessel type functions}\label{sbessel}%

We present some formulae for Bessel type functions used in this paper.
\begin{eqnarray}
\label{UF.1}
\frac{d}{dx}\left[\frac{z_n(x)}{x^n}\right]=-\frac{z_{n+1}(x)}{x^n},\qquad 
\frac{d}{dx}\left[x^{n+1}z_n(x)\right]=x^{n+1}z_{n-1}(x),
\end{eqnarray} 
where $z_n(x)$ can be spherical Bessel functions, spherical Neumann
functions, Bessel functions, and Neumann functions.

Spherical Bessel functions and spherical Neumann functions are related by
\begin{eqnarray}
y_n(x)=(-1)^{n+1}j_{-n-1}(x).
\end{eqnarray}
Their asymptotic forms are
\begin{eqnarray}
\label{UF.2}
j_n(x)\approx\frac{\sin{(x-n\pi/2)}}{x},\qquad y_n(x)\approx -\frac{\cos{(x-n\pi/2)}}{x}
\end{eqnarray}
for
$x\gg1$. If n is even, $j_n(x)\approx \pm j_0(x)$ and $y_n(x)\approx \pm y_0(x)$. If n is odd, $j_n(x)\approx \pm y_0(x)$ and $y_n(x)\approx \pm j_0(x)$.
The first and second kinds of spherical Hankel functions are defined as
\begin{eqnarray}
\label{UF.3}
h_n^{(1)}(x)=j_n(x)+iy_n(x),\qquad h_n^{(2)}(x)=j_n(x)-iy_n(x).
\end{eqnarray}
Using elementary functions, we have
\begin{eqnarray}
\label{UF.4}
j_0(x)&=&\frac{\sin{x}}{x},\\
j_1(x)&=&\frac{1}{x}\left[\frac{\sin{x}}{x}-\cos{x}\right],\\
j_2(x)&=&\frac{1}{x}\left[\left(\frac{3}{x^2}-1\right)\sin{x}-\frac{3}{x}\cos{x}\right],\\
y_0(x)&=&-\frac{\cos{x}}{x},\\
y_1(x)&=&-\frac{1}{x}\left[\frac{1}{x}\cos{x}+\sin{x}\right],\\
y_2(x)&=&-\frac{1}{x}\left[\left(\frac{3}{x^2}-1\right)\cos{x}+\frac{3}{x}\sin{x}\right],\\
h^{(1)}_1(x)&=&-\frac{1}{x}\left(1+\frac{i}{x}\right)e^{-ix},\\
h^{(2)}_1(x)&=&-\frac{1}{x}\left(1-\frac{i}{x}\right)e^{-ix}.
\end{eqnarray}

\section{Analytical solutions of wave equation}\label{solution}%

In this Appendix we shall discuss solutions of  the equation of motion
[Eq.~(\ref{2.8})]. 
While we assume $\Pi_{ij}=0$ in this Appendix, we shall treat
$\Pi_{ij}\neq 0$ in Appendix \ref{neutrinodamp}.
Imposing appropriate boundary conditions \cite{QFC}, one obtains simple
analytical solutions for tensor modes of fluctuations in the
inflationary (de Sitter), radiation dominated (RD) and matter dominated
(MD) universe, as
\begin{eqnarray}
\label{2.0.1a}
h_{\mathbf{k}}(\tau)&=&\frac{\sqrt{16\pi
G}}{\sqrt{2k}a}\bigg(1-\frac{i}{k\tau}\bigg)e^{-ik\tau}\alpha(\mathbf{k}),\nonumber\\
&=&-\frac{\tau}{a}\sqrt{8\pi Gk}h^{(2)}_1\!(k\tau)\alpha(\mathbf{k})\qquad\,\textrm{inflation},\\
\label{2.0.1b}
h_{\mathbf{k}}(\tau)&=&\left[j_0(k\tau)\right]h_{\mathbf{k}}^{prim} \qquad \qquad\qquad\;\;\textrm{RD}, \\
\label{2.0.1c}
h_{\mathbf{k}}(\tau)&=&\left[ \frac{3j_1(k\tau)}{k\tau}\right] h_{\mathbf{k}}^{prim} \qquad\qquad\quad \textrm{MD},
\end{eqnarray}
where $\alpha(\mathbf{k})$ is a stochastic variable satisfying $\langle\alpha(\mathbf{k})\alpha^*(\mathbf{k}')\rangle=\delta^3(\mathbf{k}-\mathbf{k}')$, and spherical Bessel-type functions are given in Appendix \ref{sbessel}. 
We classify wave modes by their horizon crossing time, $\tau_{\rm{hc}}$;
\begin{eqnarray}
\label{2.0.1d}
\mid\mathbf{k}\mid=k\left\{
      \begin{array}{c}
       >k_{\rm{eq}} \quad \textrm{the modes that entered the horizon
	during RD: } \tau_{\rm{hc}}<\tau_{\rm{eq}}\\
       <k_{\rm{eq}} \quad \textrm{the modes that entered the horizon
	during MD: } \tau_{\rm{hc}}>\tau_{\rm{eq}}
      \end{array}\right.,
\end{eqnarray} 
where $\tau_{\rm{eq}}$ denotes the time at the matter-radiation
equality, and $\tau_{\rm{hc}}$ denotes the time when
fluctuation modes crossed the horizon, $k\tau_{\rm{hc}}= 1$.
Notice that $|h_k(\tau)|^2$ for each solution
(\ref{2.0.1a}) - (\ref{2.0.1c}) does not depend on time ($\equiv|h_k^{prim}|^2$) at the super-horizon scale, $|k\tau|\ll1$.

The tensor mode fluctuations from the inflationary universe left the
horizon and \textit{froze out}. 
Its dimensionless spectrum is given from Eq. (\ref{2.0.1a}) as
\begin{eqnarray}
\label{2.0.2}
\Delta_h^2(k)&\equiv& 4k^3\frac{|h_k^{\rm{inf}}|^2}{2\pi^2}
=64\pi G\left(\frac{H_{\rm{inf}}k\tau}{2\pi}\right)^2\left(1+\frac{1}{k^2\tau^2}\right)\nonumber\\
&\simeq& \frac{16}{\pi}\left(\frac{H_{\rm{inf}}}{m_{\rm{Pl}}}\right)^2
\equiv 4k^3\frac{|h_k^{prim}|^2}{2\pi^2} \qquad(|k\tau|\ll 1),
\end{eqnarray}  
where $H_{\mathrm{inf}}$ is the Hubble parameter during inflation and $\tau=-1/(aH_{\mathrm{inf}})$ is used in the second equality.
Note that the conventional factor 4 is from $\int\frac{dk}{k}\Delta^2_h(k)\equiv \langle h_{ij}h^{ij}\rangle =2[\langle |h_{+}|^2\rangle +\langle |h_{\times}|^2\rangle]=4|h|^2$, where $|h_{+,k}|=|h_{\times,k}|\equiv|h|$ is assumed \cite{WMAP}.
From the Friedman equation during inflation, one obtains
$H^2_{\rm{inf}}\approx\frac{8\pi}{3m^2_{\rm{Pl}}}V(\phi)$, which gives 
$\Delta_{h,prim}^2\approx10V(\phi)/m^4_{\rm{Pl}}$; 
thus $\Delta^2_{h,prim}$ is sensitive to the shape of inflaton potential \cite{TurnerWL, Turner}. 
The dimensionless spectrum (\ref{2.0.2}) is nearly independent $k$. 
This is the famous prediction of the inflationary scenario known as a nearly  scale invariant spectrum.
As long as we consider de Sitter inflation, the spectrum is exactly
scale invariant, i.e. $\propto k^0$ as $\phi$ is at rest.

\begin{figure}[htb]
  \begin{center}
    \resizebox{100mm}{!}{\includegraphics{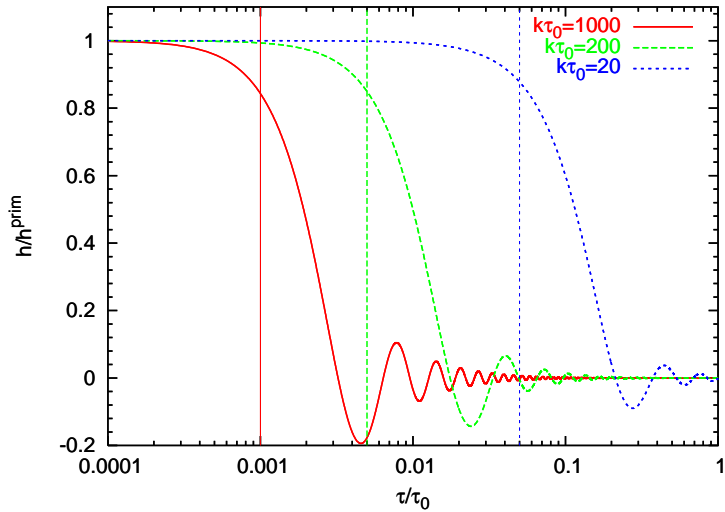}}
    \caption{Numerical solutions of tensor perturbations. The solid,
   dashed, and short-dashed lines show the high, medium, and low
   frequency modes, respectively. The higher $k$-modes enter the horizon
   earlier, and are damped more by the cosmological redshift. Vertical lines define the horizon crossing time for each $k$-mode.}
    \label{solution3}
  \end{center}
\end{figure}

\begin{figure}[htb]
  \begin{center}
    \resizebox{100mm}{!}{\includegraphics{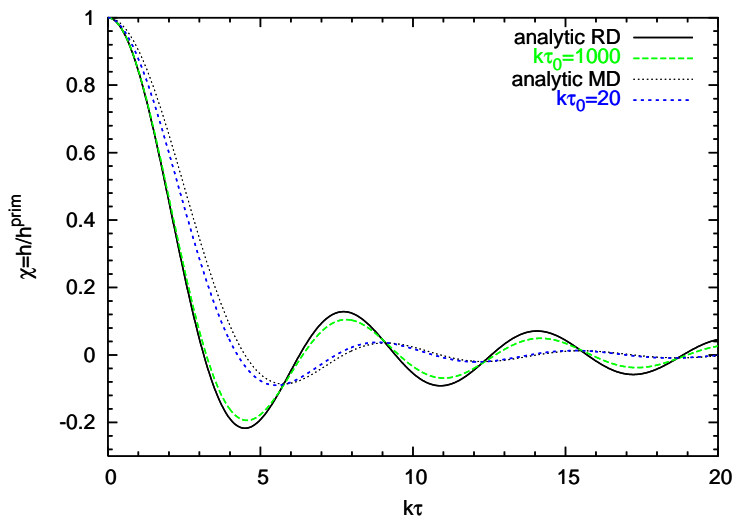}}
    \caption{Comparison between numerical solutions and analytical
   solutions of tensor perturbations. The dashed and short-dashed lines show
   numerical solutions of the high and low frequency modes,
   respectively. The higher $k$-modes enter the horizon earlier, and
   thus the numerical solution is well approximated by the analytical
   solution during the radiation era, $\chi(k\tau)=j_0(k\tau)$ (solid
   line). On the other hand, the lower $k$-modes enter the horizon
   much later, and thus the numerical solution is close to the
   analytical solution during the matter era, $\chi(k\tau)=3j_1(k\tau)/k\tau$ (dotted line).}
    \label{nodampsolution2}
  \end{center}
\end{figure}

Using the transfer function [Eq.~(\ref{2.0.2a})], we obtain the time
evolution of the amplitude of gravitational waves as 
\begin{eqnarray}
\label{2.0.2b}
\mathcal{T}(\tau<\tau_{\rm{eq}},k>k_{\rm{eq}})&=&j_0(k\tau),\\
\label{2.0.2bb}
\mathcal{T}(\tau>\tau_{\rm{eq}},k>k_{\rm{eq}})&=&\frac{\tau_{eq}}{\tau}\left[A(k)j_1(k\tau)+B(k)y_1(k\tau)\right],\\
\label{2.0.2c}
\mathcal{T}(\tau,k<k_{\rm{eq}})&=&\frac{3j_1(k\tau)}{k\tau},
\end{eqnarray} 
where 
\begin{eqnarray}
\label{2.0.2d}
A(k)&=&\frac{3}{2k\tau_{\mathrm{eq}}}-\frac{\cos{2k\tau_{\mathrm{eq}}}}{2k\tau_{\mathrm{eq}}}+\frac{\sin{2k\tau_{\mathrm{eq}}}}{(k\tau_{\mathrm{eq}})^2},\\
\label{2.0.2e}
B(k)&=&-1+\frac{1}{(k\tau_{\mathrm{eq}})^2}-\frac{\cos{2k\tau_{\mathrm{eq}}}}{(k\tau_{\mathrm{eq}})^2}-\frac{\sin{2k\tau_{\mathrm{eq}}}}{2k\tau_{\mathrm{eq}}}.
\end{eqnarray}
Their conformal time derivatives are given as
\begin{eqnarray}
\label{2.0.2pb}
\mathcal{T'}(\tau<\tau_{\rm{eq}},k>k_{\rm{eq}})&=&-kj_1(k\tau),\\
\label{2.0.2pbb}
\mathcal{T'}(\tau>\tau_{\rm{eq}},k>k_{\rm{eq}})&=&-\frac{k\tau_{eq}}{\tau}\left[A(k)j_2(k\tau)+B(k)y_2(k\tau)\right],\\
\label{2.0.2pc}
\mathcal{T'}(\tau,k<k_{\rm{eq}})&=&-\frac{3j_2(k\tau)}{\tau}.
\end{eqnarray} 
Eqs. (\ref{2.0.2b}) and (\ref{2.0.2bb}) are the evolution of modes which entered the horizon during the radiation era, while Eq. (\ref{2.0.2c}) is the evolution of modes which entered the horizon during the matter era.
Coefficients $A(k)$ and $B(k)$ are obtained by equating
a solution (\ref{2.0.2b}) with (\ref{2.0.2bb}) and their first derivatives [(\ref{2.0.2pb}) and (\ref{2.0.2pbb})] at the matter-radiation equality. 
The transfer function for the intermediate regime, Eq.~(\ref{2.0.2bb}),
can be calculated numerically so that the two other limiting solutions
match smoothly (See Fig.~\ref{fig1}).
If the wavelength of the gravitational waves is much shorter than the duration of the
cosmological transition, a WKB approximation may be appropriate \cite{Pritchard, Ng}.
Here we just assumed the instantaneous transition to illustrate the main point.
The analytical solutions as well as numerical solutions are presented
and compared in Fig.~\ref{solution3} and \ref{nodampsolution2}. 
The higher $k$-modes enter the horizon earlier, and their amplitudes are damped more by the cosmological redshift.

\section{The relative spectral density: $\Omega_h(k)$}\label{omega}%

In this Appendix we shall define the energy momentum tensor of gravitational waves following the argument and the definition in
\S35.7 and \S35.13 of \cite{MTW}. The Ricci tensor for the metric of the
form given in Eq.~(\ref{2.1}) may be expanded in metric perturbations,
$h$:
\begin{eqnarray}
\label{2.1.01}
R_{\mu\nu}=\bar{R}_{\mu\nu}+R^{(1)}_{\mu\nu}+R^{(2)}_{\mu\nu}+\mathcal{O}(h^3),
\end{eqnarray}
where $R^{(1)}_{\mu\nu}\sim \mathcal{O}(h)$ and $R^{(2)}_{\mu\nu}\sim
 \mathcal{O}(h^2)$.

For the vacuum field equation, $R_{\mu\nu}=0$. 
As the Einstein equation is non-linear, 
$\bar{R}_{\mu\nu}$ is in general not linear in $h_{\mu\nu}$. 
The linear term in Eq.~(\ref{2.1.01}) must obey the vacuum equation,
\begin{eqnarray}
\label{2.1.02}
R^{(1)}_{\mu\nu}=0.
\end{eqnarray}
 This is an equation for the propagation of the gravitational waves,
 which corresponds to Eq. (\ref{2.8}) or more generally to
 Eq.~(\ref{4.1}) in the FRW universe. The remaining part of $R_{\mu\nu}$
 may be divided into a smooth part which varies only on scales larger
 than some coarse-graining scales,
\begin{eqnarray}
\label{2.1.03}
\bar{R}_{\mu\nu}+\langle R^{(2)}_{\mu\nu}\rangle=0,
\end{eqnarray}
and a fluctuating part which varies on smaller scales
\begin{eqnarray}
\label{2.1.03b}
R^{(1)\rm{nonlinear}}_{\mu\nu}+R^{(2)}_{\mu\nu}-\langle R^{(2)}_{\mu\nu}\rangle=0,
\end{eqnarray}
up to the second order in $h_{\mu\nu}$. Here, $R^{(1)\rm{nonlinear}}_{\mu\nu}$ is defined by Eq.~(\ref{2.1.03b}) and represents the nonlinear correction to the propagation of $h_{\mu\nu}$, Eq.~(\ref{2.1.02}), which gives $h_{\mu\nu}\to h_{\mu\nu}+j_{\mu\nu}$, where $j_{\mu\nu}\sim \mathcal{O}(h^2)$ \cite{MTW}. 
Eq.~(\ref{2.1.03}) represents how the stress energy in the gravitational waves creates the background curvature. The Einstein equation in vacuum is then
\begin{eqnarray}
\label{2.1.04}
\bar{G}_{\mu\nu}=\bar{R}_{\mu\nu}-\frac{1}{2}\bar{R}\bar{g}_{\mu\nu}=8\pi GT^{(GW)}_{\mu\nu},\\
\label{2.1.04b}
T^{(GW)}_{\mu\nu}\equiv -\frac{1}{8\pi G}\left(\langle R^{(2)}_{\mu\nu}\rangle-\frac{1}{2}\bar{g}_{\mu\nu}\langle R^{(2)}\rangle\right),
\end{eqnarray}
where $T^{(GW)}_{\mu\nu}$ is a definition of the energy momentum tensor for the gravitational waves and $\langle\quad \rangle$ denotes an average over several wavelengths. 
The importance of the effective energy momentum tensor is that it tells
us 
how \textit{backreaction} from energy density of gravitational waves would affect the expansion law of the background universe. Note that the \textit{effective} energy momentum tensor defined by Eq. (\ref{2.1.04b}) is different from that defined by the Neother current of the Lagrangian density, $T^{\rm{Neother}}_{\mu\nu}\equiv \frac{2}{\sqrt{-g}}\frac{\delta S^{(2)}}{\delta g^{\mu\nu}}$, where $S^{(2)}$ is the second order perturbation in the Einstein-Hilbert action.
These definitions coincide only deep inside the horizon. 
Note also that in the notation of \cite{MTW}, $G=1$, but in our notation, $\hbar =c=1,\,G=m_{\rm{Pl}}^{-2}$,where $m_{\rm{Pl}}$ is the Plank mass.
Since $\langle R^{(2)}\rangle=0$ \cite{MTW},
\begin{eqnarray}
\label{2.1.05}
T^{(GW)}_{\mu\nu}=\frac{1}{32\pi G}\langle h_{\alpha\beta|\mu}h^{\alpha\beta}{}_{|\nu}\rangle = \frac{1}{32\pi G}\langle h_{\alpha\beta,\mu}h^{\alpha\beta}{}_{,\nu}\rangle +\mathcal{O}(h^3),
\end{eqnarray}
where $|$ is the covariant derivative with respective to background metric, $\bar{g}_{\mu\nu}$. Note that we have employed the transverse-traceless (TT) gauge. In linear theory we neglect higher order terms in the energy momentum tensor.

The energy density of gravitational waves, $\rho_h$, is defined by the 0-0
component of the energy momentum tensor.
\begin{eqnarray}
\label{2.1.1}
\rho_h(\tau) \equiv T^{(GW)}_{00}=\frac{1}{32\pi G}\langle \dot{h}_{ij}\dot{h}^{ij}\rangle,
\end{eqnarray}
where $h_{ij}$ is in the TT gauge.
There are only two independent modes for gravitational waves;
\begin{eqnarray}
\label{2.1.2}
h_{ij}=\left(
           \begin{array}{ccc}
           h_{+}     & h_{\times} & 0 \\
           h_{\times}&-h_+        & 0 \\
           0         & 0          & 0
           \end{array}\right), 
\end{eqnarray}
where + and $\times$ denote two independent polarization modes and their
propagation direction is taken in $\hat{z}$ direction. Hence,
\begin{eqnarray}
\label{2.1.3}
\rho_h(\tau) &=&\frac{2}{32\pi G}\langle \dot{h}_+^2+\dot{h}_{\times}^2\rangle
\nonumber\\
&=& \frac{1}{16\pi Ga^2}\langle {h'}_+^2 +{h'}_{\times}^2\rangle
\nonumber\\
&=& \frac{1}{16\pi Ga^2}\int\frac{d^3\!k}{(2\pi)^3} \int\frac{d^3\!k'}{(2\pi)^3}
\langle\left( {h'}_{+,\mathbf{k}}{h'}_{+,\mathbf{k}'} + {h'}_{\times,\mathbf{k}}{h'}_{\times,\mathbf{k}'}\right) e^{i(\mathbf{k}+\mathbf{k}')\cdot\mathbf{x}}\rangle,
\end{eqnarray}
where Fourier transformation was done and ${h^*}_{\lambda,\mathbf{k}}={h}_{\lambda,-\mathbf{k}}$ in the last step. For stochastic
modes, the spatial average over several wavelengths, $\langle\quad\rangle$, is equivalent
to the ensemble average in k space;
 \begin{eqnarray}
\label{2.1.4}
\langle {h'}_{\lambda,\mathbf{k}}{h'}_{\lambda ',\mathbf{k}'}\rangle=
(2\pi)^3\delta_{\lambda,\lambda '}\delta^{(3)}(\mathbf{k}+\mathbf{k}')| h_{\lambda,k}'|^2,
\end{eqnarray}
where $\lambda=+,\times$. Using (\ref{2.1.3}) and (\ref{2.1.4}), we obtain
\begin{eqnarray}
\label{2.1.5}
\rho_h(\tau) = \frac{1}{16\pi Ga^2}\int\frac{d^3\!k}{(2\pi)^3}\left[
|{h'}_{+,k}(\tau)|^2 +  |{h'}_{\times,k}(\tau)|^2\right].
\end{eqnarray}

It is reasonable to assume that the primordial gravitational waves are
unpolarized, i.e. $|h_{+,k}|^2=|h_{\times,k}|^2$.
Whenever we express the time evolution of some quantities, it is convenient to express them in terms of the transfer function,$\mathcal{T}(k\tau)$, and the primordial amplitude,$\Delta_{h,prim}^2$, defined as (\ref{2.0.2a});
\begin{eqnarray}
\label{2.1.6}
\rho_h(\tau) = \frac{1}{32\pi Ga^2}\int d\ln
k\Delta_{h,prim}^2\left[\mathcal{T}'(k\tau)\right]^2,
\end{eqnarray}
where
\begin{eqnarray}
\label{2.1.7}
\Delta_{h,prim}^2\equiv4\frac{k^3}{2\pi^2}|h^{prim}_{k}|^2=\frac{16}{\pi}\left(\frac{H_{\mathrm{inf}}}{m_{Pl}}\right)^2.
\end{eqnarray}
Here, $|h_k^{prim}|^2$ is the amplitude of gravitational waves outside the horizon, $\mid
k\tau\mid\ll 1$, during inflation. Well inside the horizon averaging over several periods, the leading term of
$\overline{[\mathcal{T}'(k\tau)]^2}$ is proportional to
$\tau^{-2}\propto a^{-2}$ during the radiation era and $\propto \tau^{-4}\propto a^{-2}$
during the matter era.
Thus $\rho_h\propto a^{-4}$, which is consistent with the fact that
graviton is massless and thus relativistic.

It is common to define the relative spectral density as the normalized
energy density per logarithmic scale.
\begin{eqnarray}
\label{2.1.9}
\Omega_h(\tau,k) &\equiv&
\frac{\tilde{\rho}_h(\tau,k)}{\rho_{\rm{cr}}(\tau)},\\
\tilde{\rho}_h(\tau,k)&\equiv& \frac{d\rho_h(\tau)}{d\ln k}, \nonumber
\end{eqnarray}
where $\rho_{\rm{cr}}(\tau)$ is critical density
of the universe, and $\tilde{\rho}_h(\tau,k)$ denotes energy density of the
gravitational waves per logarithmic scale. Inserting (\ref{2.1.6})
into (\ref{2.1.9}), we obtain
\begin{eqnarray}
\label{2.1.10}
\Omega_h(\tau,k)&=&\frac{\Delta_{h,prim}^2}{32\pi
Ga^2\rho_c(\tau)}\left[\mathcal{T}'(\tau,k)\right]^2.
\end{eqnarray}
Recalling Friedman equation, $H^2=8\pi G\rho_c/3$, (\ref{2.1.10}) becomes
\begin{eqnarray}
\label{2.1.10c}
\Omega_h(\tau,k)&=&\frac{\Delta_{h,prim}^2}{12H^2(\tau)a^2}\left[\mathcal{T}'(\tau,k)\right]^2.
\end{eqnarray}
In this paper, we shall evaluate this quantity exactly within the Standard Model of elementary particles. For an analytical model, $\mathcal{T}'(\tau,k)$ is given by Eqs.~(\ref{2.0.2pb}) -- (\ref{2.0.2pc}).

\section{Collisionless damping due to neutrino free-streaming}\label{neutrinodamp}

In this Appendix, we review the effect of collisionless particles on gravitational waves.
Treating relativistic neutrino gas by classical kinetic theory, the linearized Einstein-Boltzmann equation (\ref{2.4}) can be written as an
integro-differential equation (\ref{4.1}). The derivation of this
integro-differential equation is given in the literature, for instance
\cite{Vishniac82, Kasai&Tomita85, Rebhan&Schwarz94, Weinberg04} for both scalar and tensor modes, \cite{Bond&Szalay83} for scalar modes, and will be reviewed briefly in this Appendix.

At the temperature of $\sim 2$ MeV, where neutrinos decoupled and became out of equilibrium with photons, electrons, or positrons, the number of effective relativistic species is $g_*(\sim 2 MeV)=10.75$.\footnote{We have assumed instantaneous decoupling of neutrinos, but this is not true in general.} The free-streaming neutrino gas after their decoupling satisfies the collisionless Boltzmann equation, i.e. the Vlasov equation,
\begin{eqnarray}
\label{4.0.1}
\frac{dF(x,P)}{dt}=0,
\end{eqnarray}
where $F(x,P)=\bar{F}(P)+\delta F(x,P)$ is a distribution function. The distribution function of relativistic neutrinos is given by 
\begin{eqnarray}
\label{4.0.2}
\bar{F}(P^0)=\frac{g_{\nu}}{e^{P^0/T}+1},
\end{eqnarray}
where $g_{\nu}$ denotes the number of helicity states for neutrinos and
anti-neutrinos. Here,
$P^{\mu}\equiv \frac{dx^{\mu}}{d\lambda}$ and $P^0= \sqrt{g_{ij}P^iP^j}$, which is implied by the constraint for relativistic particles;
\begin{eqnarray}
\label{4.0.3}
g_{\mu\nu}P^{\mu}P^{\nu}=0.
\end{eqnarray}
Therefore, there are only three independent components of the momentum
vector. One can also relate $P^i$ with $P_0=-P^0$ as
\begin{eqnarray}
\label{4.0.4}
P^i=\pm\frac{\gamma^i P_0}{a}\left(1-\frac{1}{2}h_{jk}\gamma^j\gamma^k\right),
\end{eqnarray}
where $\gamma^i=\gamma_i$'s are directional cosines and $P^0$ is the energy of neutrinos.
We chose positive sign convention for $P^0\equiv \frac{dt}{d\lambda}$. 
Note that $\delta_{ij}\gamma^i\gamma^j=1$, and $P^i\equiv C\gamma^i
P_0$, where the coefficient, $C$, is obtained from Eq. (\ref{4.0.3});
\begin{eqnarray}
0&=&P_0P^0+a^2P^jP^j+a^2h_{ij}P^iP^j,\nonumber\\
0&=&-(P_0)^2+a^2C^2P_0^2+a^2h_{ij}\gamma^i\gamma^jC^2P_0^2,\nonumber\\
1&=&a^2C^2(1+h_{ij}\gamma^i\gamma^j).\nonumber 
\end{eqnarray}
We consider tensor perturbations. Eq. (\ref{4.0.1}) can be expressed as
\begin{eqnarray}
\label{4.0.5}
\frac{dF(t,x^i,\gamma^i,P^0)}{dt}=\frac{\partial F}{\partial t}+\frac{dx^i}{dt}\frac{\partial F}{\partial x^i}+ \frac{dP^0}{dt}\frac{\partial F}{\partial P^0}+\frac{d\gamma^i}{dt}\frac{\partial F}{\partial \gamma^i}=0.
\end{eqnarray}
The last term is negligible in the linear perturbation theory, as $\frac{\partial F}{\partial \gamma^i}$ is of the first order in perturbations and $\dot{\gamma^i}=-\frac{1}{2a}h_{jk,i}\gamma^j\gamma^k$.

For the second term $\frac{\partial F}{\partial x^i}$ is of the first order in perturbations and
\begin{eqnarray}
\label{4.0.6}
\frac{dx^i}{dt}=\frac{dx^i}{d\lambda}\frac{d\lambda}{dt}=\frac{P^i}{P^0}.
\end{eqnarray} 
Using Eq. (\ref{4.0.4}), one obtains
\begin{eqnarray}
\label{4.0.7}
\frac{dx^i}{dt}\frac{\partial F}{\partial x^i}=\frac{\gamma^i}{a}\frac{\partial F}{\partial x^i}
\end{eqnarray} 
in the leading order, as $\bar{F}$ does not depend on $x^i$; thus, $\frac{\partial F}{\partial x^i}$ is a perturbation.

\begin{figure}[htb]
  \begin{center}
    \resizebox{100mm}{!}{\includegraphics{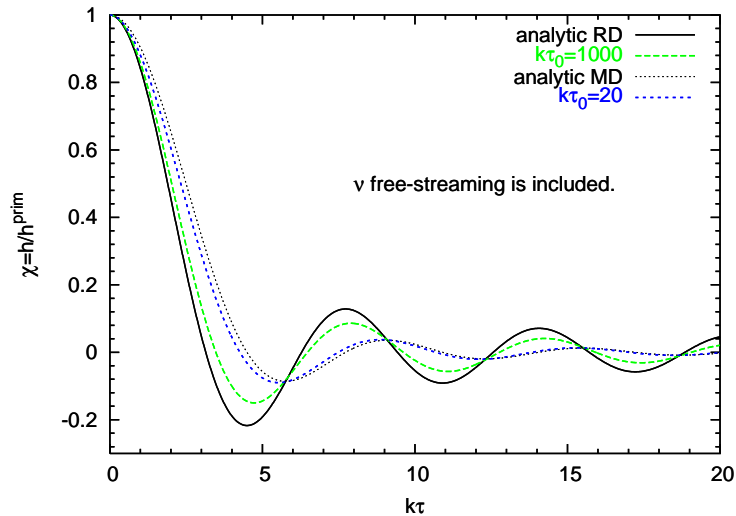}}
    \caption{Comparison between numerical solutions and analytical
   solutions of tensor perturbations. The effect of neutrino free-streaming is included for numerical solutions, but not for analytical solutions. The dashed and short-dashed lines show
   numerical solutions of the high and low frequency modes,
   respectively. The higher $k$-modes enter the horizon during the radiation era after neutrino decoupling, and
   thus the numerical solution is damped by neutrino free-streaming compared to the analytical
   solution, $\chi(k\tau)=j_0(k\tau)$ (solid
   line). On the other hand, the lower $k$-modes enter the horizon
   much later, and thus the numerical solution is closer to the
   analytical solution during the matter era, $\chi(k\tau)=3j_1(k\tau)/k\tau$ (dotted line).}
    \label{solution2}
  \end{center}
\end{figure}

\begin{figure}[htb]
  \begin{center}
    \resizebox{100mm}{!}{\includegraphics{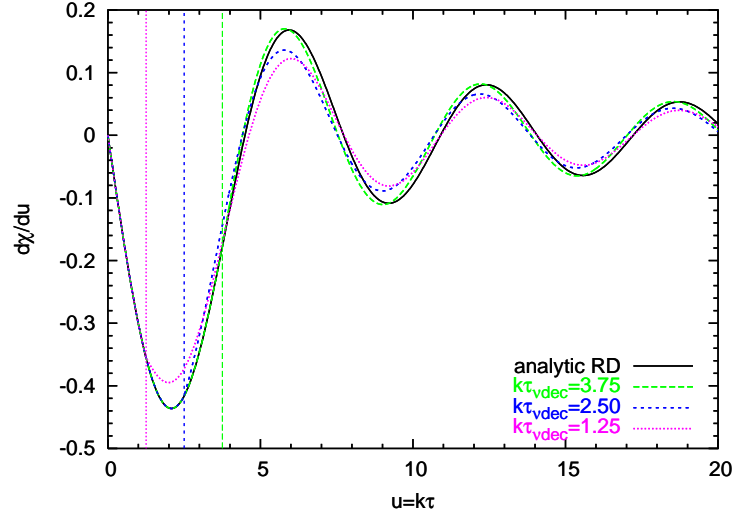}}
    \caption{Derivatives of modes which entered the horizon before
   neutrino decoupling. The solid line shows an analytical solution,
   $\chi'=-j_1(u)$, during the radiation era without neutrino
   decoupling. The dotted, short-dashed, and dashed lines show numerical
   solutions of $\chi'(k\tau)$ for which neutrinos decoupled at
   $\tau_{\nu\rm dec}$ given by $k\tau_{\nu\rm{dec}}=1.25,\;2.5,\;3.75$, respectively. They are damped by giving energy to free-streaming neutrinos. Vertical lines indicate the neutrino decoupling time for each mode.}
    \label{solutionRD1p}
  \end{center}
\end{figure}

\begin{figure}[htb]
  \begin{center}
    \resizebox{100mm}{!}{\includegraphics{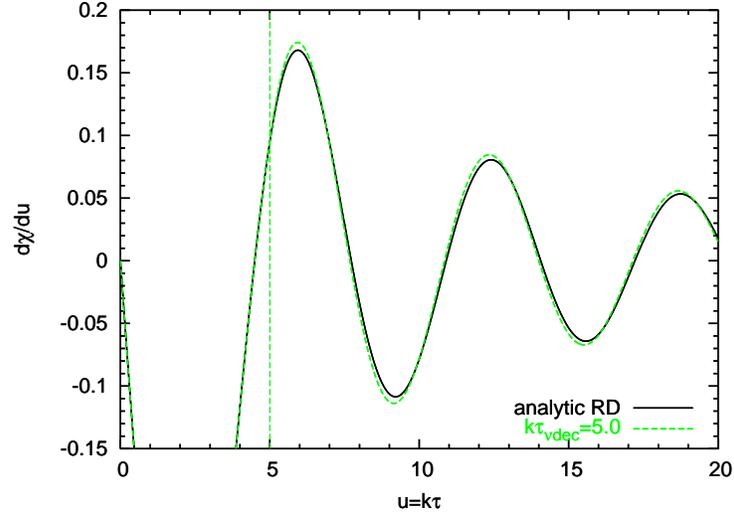}}
    \caption{Derivative of a mode which entered the horizon before neutrino decoupling. The solid line shows an analytic solution,$\chi'=-j_1(u)$, during the radiation era without neutrino decoupling. The dashed line shows numerical solutions of $\chi'(k\tau)$ for which neutrinos decoupled at
   $\tau_{\nu\rm dec}$ given by $k\tau_{\nu\rm{dec}}=5.0$. The wave is amplified by gaining energy from free-streaming neutrinos. The vertical line indicates the neutrino decoupling time.}
    \label{solutionRD2p}
  \end{center}
\end{figure}

 For the third term we use the geodesic equation,
\begin{eqnarray}
\label{4.0.8}
\frac{dP^{\mu}}{d\lambda}&=&-\Gamma^{\mu}{}_{\alpha\beta}P^{\alpha}P^{\beta},\\
\Gamma^{\mu}{}_{\alpha\beta}&=&\frac{g^{\mu\nu}}{2}\left[\frac{\partial g_{\alpha\nu}}{\partial x^{\beta}}+\frac{\partial g_{\beta\nu}}{\partial x^{\alpha}}-\frac{\partial g_{\alpha\beta}}{\partial x^{\nu}}\right].
\end{eqnarray} 
The time component of the geodesic equation is
\begin{eqnarray}
\label{4.0.9}
\frac{dt}{d\lambda}\frac{dP^0}{dt}&=&-\Gamma^{0}{}_{\alpha\beta}P^{\alpha}P^{\beta},\nonumber\\
&=&-\frac{g^{0\nu}}{2}\left[2\frac{\partial g_{\alpha\nu}}{\partial x^{\beta}}-\frac{\partial g_{\alpha\beta}}{\partial x^{\nu}}\right]P^{\alpha}P^{\beta},\nonumber\\
&=&-\frac{\dot{a}}{a}(P^0)^2-\frac{1}{2}a^2\frac{\partial h_{ij}}{\partial t}P^iP^j,
\end{eqnarray} 
where $g_{00}=-1,\;g_{0i}=0$ were used from the second line to the last line.
Up to the first order in perturbations
\begin{eqnarray}
\label{4.0.10}
\frac{1}{P^0}\frac{dP^0}{dt}
=-\frac{\dot{a}}{a}-\frac{1}{2}\frac{\partial h_{ij}}{\partial t}\gamma^i\gamma^j,
\end{eqnarray} 
where we have used Eq. (\ref{4.0.4}) and neglected higher order terms. This equation describes the change in the neutrino energy as it propagates in a FRW universe with gravitational waves. The first term accounts for the redshift of energy due to an isotropic expansion. The second term tells us that neutrinos lose energy if $\frac{\partial h_{ij}}{\partial t}>0$, or gain energy if $\frac{\partial h_{ij}}{\partial t}<0$ from gravitational waves. This energy flow from neutrinos to gravitational waves causes collisionless damping (Figs.~\ref{solution2} and \ref{solutionRD1p}) and amplification (Fig.~\ref{solutionRD2p}) of gravitational waves.

Finally, by combining Eqs. (\ref{4.0.5}), (\ref{4.0.7}), and (\ref{4.0.10}), the Vlasov equation for the first order perturbations is obtained as
\begin{eqnarray}
\label{4.0.11}
\left(\frac{dF}{dt}\right)_{\rm{first\;order}}
=\frac{\partial \delta F}{\partial t}
+\frac{\gamma^i}{a}\frac{\partial \delta F}{\partial x^i}
-P^0\frac{\partial \delta F}{\partial P^0}\frac{\dot{a}}{a}
-P^0\frac{\partial \bar{F}}{\partial P^0}\frac{1}{2}\frac{\partial h_{ij}}{\partial t}\gamma^i\gamma^j=0,
\end{eqnarray}
where $F=\bar{F}+\delta F(t,x^i,\gamma^i,P^0)$ and $\delta F$ is a tensor type perturbation in a distribution function of neutrinos. 
The zeroth order Vlasov equation merely gives cosmological redshift, $P^0\propto a^{-1}$, as explained above. 
Defining $\mu\equiv \gamma^i k_i/k$ and Fourier transforming Eq. (\ref{4.0.11}), the first order Vlasov equation in the momentum space is given as
 \begin{eqnarray}
\label{4.0.12}
\frac{\partial f_k}{\partial t}-\frac{\dot{a}}{a}P^0\frac{\partial f_k}{\partial P^0}+\frac{ik\mu}{a}f_k=P^0\frac{\partial \bar{F}}{\partial P^0}\frac{1}{2}\frac{\partial h_{k}}{\partial t},
\end{eqnarray}
where we have used 
\begin{eqnarray}
h_{ij}(t,\mathbf{x})&=&\sum_{\lambda=+,\times}\int\!\!\frac{d^3\!k}{(2\pi)^3}h_{\lambda,k}(t)Q_{ij}^{\lambda}(\mathbf{x}),\\
\delta F&=& \sum_{\lambda=+,\times}\int\!\!\frac{d^3\!k}{(2\pi)^3}f_{\lambda,k}(t,P^0,\mu)\gamma^i\gamma^jQ_{ij}^{\lambda}(\mathbf{x}).
\end{eqnarray}
Here, tensor harmonics $Q_{ij}^{\lambda}(\mathbf{x})$ are solutions of the tensor Helmholtz equation; $Q_{ij|a}^{\lambda}{}^{|a}(\mathbf{x})+k^2Q_{ij}^{\lambda}(\mathbf{x})=0$, $\partial_l Q_{ij}^{\lambda}=ik_l Q_{ij}^{\lambda}$. 
They are symmetric, traceless, and divergenceless; $Q_{ij}^{\lambda}=Q_{ji}^{\lambda}$, $\gamma^{ij}Q_{ij}^{\lambda}=Q_{ij}^{\lambda}{}^{|j}=0$, where $\gamma^{ij}\equiv a^{2}\bar{g}^{ij}$ and $|$ denotes the covariant derivative with respect to the spatial metric $\gamma^{ij}$. 
Note that Fourier transformation here is the generalization of Eq. (\ref{2.3}) for arbitrary spatial geometry of the universe. One can treat $Q_{ij}^{\lambda}(\mathbf{x})$ as a plane wave in a flat geometry case.

Due to the existence of the second term on the left-hand side of Eq. (\ref{4.0.12}), we cannot solve this equation. 
Thus following \cite{Bond&Szalay83}, we introduce the comoving momentum,
$q^{\mu}\equiv aP^{\mu}$. Regarding $F$ as a function of comoving
energy, $q\equiv q^0$, and conformal time, $\tau$, the third term in
Eq. (\ref{4.0.5}) may be replaced by $\frac{dq}{d\tau}\frac{\partial
F}{\partial q}=-\frac{1}{2}qh'_{ij}\gamma^i\gamma^j\frac{\partial
\bar{F}}{\partial q}$ up to the linear order.
Then the linearized Vlasov equation, $\frac{d}{d\tau}F(\tau,x^i,\gamma^i,q)=0$, becomes
\begin{eqnarray}
\label{4.0.12b}
\frac{\partial f_k}{\partial \tau}+ik\mu f_k=q\frac{\partial \bar{F}}{\partial q}\frac{1}{2}\frac{\partial h_{k}}{\partial \tau},
\end{eqnarray}
where $f_k=f_k(\tau,q,\mu)$.
One finds the solution of Eq. (\ref{4.0.12b}) as
\begin{eqnarray}
\label{4.0.13}
f_k(\tau,q,\mu)=e^{-i\mu k(\tau-\tau_{\nu\,\rm{dec}})}f_k(\tau_{\nu \rm{dec}},q,\mu)
+\frac{q}{2}\frac{\partial \bar{F}}{\partial q}\int^{\tau}_{\tau_{\nu\,\rm{dec}}}\!\!\!\!\!\!\!\!\!\!d\tau' h'_k(\tau')e^{-i\mu k(\tau-\tau')},
\end{eqnarray}
 where the prime on $h_k(\tau)$ denotes the derivative with respect to
 the conformal time. As there is no primordial tensor perturbations in the neutrino distribution function before neutrino decoupling, $f_k(\tau_{\nu \rm{dec}},q,\mu)=0$.

The right-hand side of the linearized Einstein equation includes anisotropic stress as in Eq. (\ref{2.4});
\begin{eqnarray}
\label{4.1.1}
\delta T^{(\nu)}_{ij}=a^2\sum_{\lambda=+,\times}\int\!\!\frac{d^3\!k}{(2\pi)^3}\Pi_{\lambda,k}Q^{\lambda}_{ij}(\mathbf{x}),
\end{eqnarray}
where $T^{(\nu)}_{ij}$ denotes the stress energy tensor of neutrinos. Since $T^{(\nu)}_{ij}=\frac{1}{\sqrt{-g}}\int\!\frac{d^3q}{q^0}q_i q_jF(q)$, its perturbation can be expressed as
 \begin{eqnarray}
\label{4.1.2}
\delta T^{(\nu)}_{ij}&=& a^{-4}\int\!\frac{d^3q}{q^0}\left[\bar{q}_{i}\bar{q}_{j}\delta F+(\delta q_{i}\bar{q}_{j}+\bar{q}_{i}\delta q_{j})\bar{F}\right],\\
\delta F&=&\sum_{\lambda=+,\times}\int\!\!\frac{d^3\!k}{(2\pi)^3}f_{\lambda,k}(\tau,q,\mu)\gamma^l\gamma^mQ^{\lambda}_{lm}(\mathbf{x}).\nonumber
\end{eqnarray}
The second and the third terms of (\ref{4.1.2}) cancel out in linear perturbation theory. Thus
\begin{eqnarray}
\label{4.1.2.1}
\Pi_{\lambda,k}Q^{\lambda}_{ij}(\mathbf{x})
&=&a^{-4}\int\!\frac{d^3q}{q^0}q^2\gamma^i\gamma^j\gamma^l\gamma^m f_{\lambda,k}Q^{\lambda}_{lm}(\mathbf{x}).
\end{eqnarray}
Inserting solution of the Vlasov equation (\ref{4.0.13}) into Eq. (\ref{4.1.2.1}) and using equality $\int d\Omega_{q}\gamma^i\gamma^j\gamma^l\gamma^m e^{-i\hat{\gamma}\cdot\hat{k}u}Q^{\lambda}_{lm}=\frac{1}{8}(\delta^{il}\delta^{jm}+\delta^{im}\delta^{jl})\int d\Omega_{q}e^{-i\mu u}(1-2\mu^2+\mu^4)Q^{\lambda}_{lm}$, one obtains
\begin{eqnarray}
\label{4.1.3}
\Pi_{k}&=&\frac{1}{4a^4}\int\!d^3\!q q(1-2\mu^2+\mu^4) f_k,\nonumber\\
\label{4.1.3b}
&=&-4\bar{\rho}_{\nu}(\tau)\int^{\tau}_{\tau_{\nu\,\rm{dec}}}\!\!\!\!\!\!\!\!\!\! d\tau' 
\left(\frac{j_2[k(\tau-\tau')]}{k^2(\tau-\tau')^2}\right)h'_{k}(\tau').
\end{eqnarray}
Here, $\bar{q}_{i}=aq\gamma^i$ and $\bar{q}^{i}=a^{-1}q\gamma^i$, and $\bar{\rho}_{\nu}(\tau)=a^{-4}\int d^3qq\bar{F}(q)$ is the unperturbed neutrino energy density, and a negative sign appears on the right-hand side of Eq. (\ref{4.1.3b}) because integration by parts has been done. Also, we have used the identity 
\begin{eqnarray}
\label{4.1.4}
\frac{1}{16}\int^1_{-1}d\mu(1-2\mu^2+\mu^4)e^{-i\mu u}=\frac{j_2(u)}{u^2}.
\end{eqnarray}
Note that $\frac{j_2(-u)}{(-u)^2}=\frac{j_2(u)}{u^2}$, $\int^{\infty}_{-\infty}\frac{j_2(u)}{u^2}du=\frac{\pi}{8}$, and $\lim_{u\to 0}\frac{j_2(u)}{u^2}=\frac{1}{15}$.\footnote{In the references \cite{Weinberg04, Dicus&Repko05}, $K[u]\equiv -\frac{\sin{u}}{u^3}-\frac{3\cos{u}}{u^4}+\frac{3\sin{u}}{u^5}=\frac{1}{15}\left(j_0(u)+\frac{10}{7}j_2(u)+\frac{3}{7}j_4(u)\right)$, which is the same function as our kernel, i.e. $K[u]=\frac{j_2(u)}{u^2}$.}

Then the Einstein-Vlasov equation takes a form of an integro-differential equation;
\begin{eqnarray}
\label{4.1}
h''_{k}(\tau)+\bigg[\frac{2a'(\tau)}{a(\tau)}\bigg]h'_{k}(\tau)+ k^2 h_{k}(\tau)
=-24f_{\nu}(\tau)\bigg[\frac{a'(\tau)}{a(\tau)}\bigg]^2\int^{\tau}_{\tau_{\nu\;\rm{dec}}}\!\!\!\!\!\!\!\!\!\! d\tau'
\left[\frac{j_2[k(\tau-\tau')]}{k^2(\tau-\tau')^2}\right]h'_{k}(\tau'),
\end{eqnarray}
and the fraction of the total energy density in neutrinos is
\begin{eqnarray}
\label{4.3}
f_{\nu}(\tau)&\equiv&\frac{\bar{\rho}_{\nu}(\tau)}{\bar{\rho}(\tau)}\nonumber\\
&=&\frac{\Omega_{\nu}(a_0/a)^4}{\Omega_{M}(a_0/a)^3+(\Omega_{\gamma}+\Omega_{\nu})(a_0/a)^4}=\frac{f_{\nu}(0)}{1+a(\tau)/a_{EQ}}, 
\end{eqnarray}
where
\begin{eqnarray}
\label{4.4}
f_{\nu}(0)=\frac{\Omega_{\nu}}{\Omega_{\gamma}+\Omega_{\nu}}=0.40523.
\end{eqnarray}
The integro-differential equation~(\ref{4.1}) was studied in \cite{Weinberg04, Pritchard, Dicus&Repko05, Bashinsky05} in the cosmological context. Here we shall solve this equation numerically with all the Standard Model particles participating in the cosmic thermal plasma.
Anisotropic stress, $\Pi_k$, vanishes during the matter era, as $f_{\nu}\to 0$. Therefore, the damping effect is unimportant during the matter era.

Following \cite{Weinberg04}, we write
\begin{eqnarray}
\label{4.6}
h_{\lambda}(u)\equiv h_{\lambda}(0)\chi(u),
\end{eqnarray}
which gives 
\begin{eqnarray}
\label{4.6.1}
\chi''(u)+\bigg[\frac{2a'(u)}{a}\bigg]\chi'(u)+ \chi(u)
=-24f_{\nu}(u)\bigg[\frac{a'(u)}{a}\bigg]^2\int^u_{u_{\nu\;\rm{dec}}}\!\!\!\!\!\!\!\!\!\! dU \left[\frac{j_2(u-U)}{(u-U)^2}\right]\chi'(U),
\end{eqnarray}
where $u\equiv k\tau$, and derivatives are taken with respect to
$u$. After the end of inflation,$\tau_{\rm{end}}$, the amplitude of
cosmological fluctuations is conserved until the mode re-enter the
horizon, $h_{\lambda}(0)=h_{\lambda,\mathbf{k}}(\tau_{\rm{end}})$.
Note that the right hand side of Eq.(\ref{4.6.1}) disappears on the
super horizon scales --- neutrino free-streaming affects the tensor
metric perturbation only inside the horizon. 
The initial conditions are taken to be
\begin{eqnarray}
\label{4.7}
\chi(0)=1,\quad \chi'(0)=0.
\end{eqnarray}
We solve Eq.~(\ref{4.6.1}) numerically by two steps; (i) we obtain $a(\tau)$ and $a'(\tau)$ from the Friedman equation (\ref{3.0.4}) with $g_*(\tau)$ in Sec.~\ref{gstar} [Fig.~\ref{a0}], and (ii) we solve Eq.~(\ref{4.6.1}) with the scale factor that we obtained in the step (i)
The numerical solutions as well as analytical solutions are presented and compared in Fig.~\ref{solution2}. The higher Fourier modes enter the horizon during the radiation era, but after neutrino decoupling. Thus they are damped due to the presence of the right-hand side of Eq.~(\ref{4.6.1}).

In order to estimate the damping effect, let us consider the radiation era after neutrino decoupling.
During the radiation era, $a'(u)/a=1/u$, the analytical solution is
given by $\chi(u)=j_0(u)$ in the absence of neutrino free-streaming in Eq.~(\ref{4.6.1}). 
In the presence of neutrino free-streaming, the solution becomes asymptotically ($u\gg 1$) 
\begin{eqnarray}
\label{4.8}
\chi(u)\to A\frac{\sin{(u+\delta)}}{u},
\end{eqnarray}
where $A=0.80313$ and $\delta=0$ are obtained from our numerical calculations.
This asymptotic solution is valid only for rather long wavelength modes which entered the horizon well after the neutrino decoupling. 
The suppression factor $A^2=0.64502$ applies to the gravitational
wave spectrum of the modes that entered the horizon after neutrino
decoupling but before matter domination.


In order to understand the shape of the spectrum, Figs.~\ref{fig4} and \ref{fig5}, we need to consider shorter wavelength modes as well.
This may be understood as follows. As we saw in Eq.~(\ref{4.0.10}), if the time derivative of the mode is negative (positive), the mode is damped (amplified). Integrating the amplitude of gravitational waves over time, the net effect of neutrino free-streaming almost always damps gravitational waves. This is because the contribution is mainly from the first period of $\chi'(u)$, where the first trough is larger than the first peak. 
In previous paragraph we have considered the modes with $k\tau_{\nu\rm{dec}}<1$. 
Now let us consider the higher $k$-modes with $k\tau_{\nu\rm{dec}}\sim 1$, or $k\sim 10^{-10}-10^{-9}$~Hz. Note that $k\tau_{\nu\rm{dec}}=1$ represents the mode which entered the horizon at the neutrino decoupling time, $\tau_{\nu\rm{dec}}$.
The mode with larger wavenumbers would enter the horizon earlier.
Fig.~\ref{solutionRD1p} shows numerical solutions of
$\chi'(u)$ for which neutrinos decoupled at $\tau_{\nu\rm{dec}}$ given by $k\tau_{\nu\rm{dec}}=1.25,\;2.5$, or $3.75$. 
For $k\tau_{\nu\rm{dec}}=1.25$ and $2.5$, neutrinos decoupled at the first trough of $\chi'(u)$, where $\chi'(u)$ is negative. Thus their amplitudes are damped by giving energy to free-streaming neutrinos (see Eq.~(\ref{4.0.10}) and discussion below it). For $k\tau_{\nu\rm{dec}}=3.75$, neutrinos decoupled right after the first trough of $\chi'(u)$, where $\chi'(u)$ is closer to zero. Thus its amplitude is unchanged, but its phase is delayed. 
Fig.~\ref{solutionRD2p} shows numerical solutions of
$\chi'(u)$ with $k\tau_{\nu\rm{dec}}=5.0$.
For $k\tau_{\nu\rm{dec}}=5.0$, neutrinos decoupled at the first peak of $\chi'(u)$, where $\chi'(u)$ is positive. Thus the amplitude of gravitational waves is actually amplified by gaining energy from free-streaming neutrinos, and we can see this feature on the spectrum, Fig.~\ref{fig5}, at $\sim 5\times 10^{-10}$~Hz.     
Neutrino free-streaming makes gravitational waves either damp or amplify depending on their frequencies. Note that this feature is generic to instantaneous decoupling of any kinds of particles, but not realistic for neutrinos as we mentioned in Sec.~\ref{prediction}.

For extremely short wavelength modes which have already been inside the horizon before neutrino decoupling, $k\tau_{\nu\rm{dec}}\gg 1$ or $k>10^{-9}$~Hz, the suppression becomes negligibly small; $A\to1$, but the phase delay, $\delta$, is non-zero. 
These modes are undamped as positive and negative contributions of $\chi'$ to the gravitational wave energy cancel out each other after several periods of $\chi'$. No net energy conversion from gravitational waves to free-streaming neutrinos would occur.

\section{Oscillation due to drastic change of $g_*(\tau)$}\label{oscillation}%

In this Appendix we explain the effect on the gravitational wave
spectrum from a sudden change in the number of relativistic species,
$g_*$. 
To do this, we need to calculate $\Omega_h^{\rm{out}}(k_2)/\Omega_h^{\rm{in}}(k_1)$, where $k_2\ne k_1$.
In Sec.~\ref{prediction} we have already seen the numerical prediction
of the gravitational wave spectrum. In subsection~\ref{gstar2} we
provided the way to understand the relative suppression of gravitational
waves at a given $k$ ($=k_1=k_2$) with and without changes in $g_*$.
We shall discuss in a similar way what would happen to different Fourier
modes, in order to fully understand imprints of $g_*$ on the spectrum, such as oscillations, which are from cosmological events that change $g_*$ instantaneously or drastically.

\begin{figure}
  \begin{center}
    \resizebox{100mm}{!}{\includegraphics{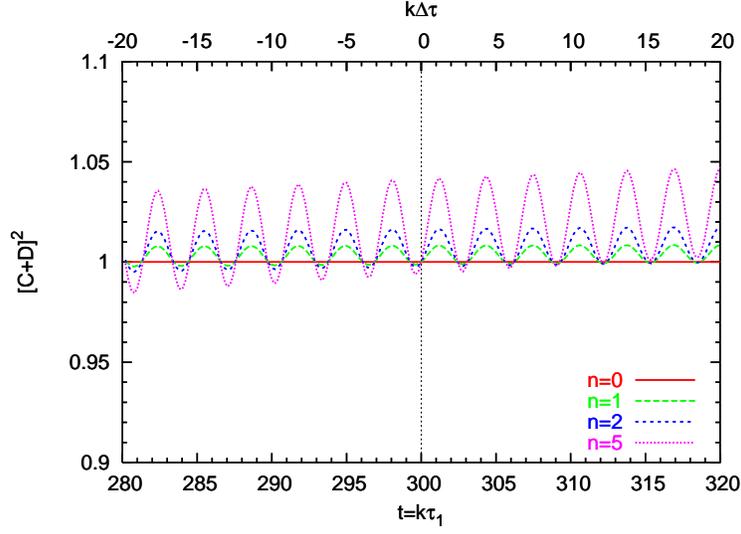}}
    \caption{The oscillatory factor $[C+D]^2$ with respect to $t\equiv k\tau_1$. The vertical line indicates $s\equiv k\tau_2=300\le t$ and $\Delta\tau\equiv\tau_1-\tau_2=0$. The solid, dashed, short-dashed, and dotted lines show $n=0, 1, 2$, and $5$ respectively. The factor, $[C+D]^2$, takes on unity at $\Delta \tau \to0$ regardless of $n$.}
    \label{CplusD2}
  \end{center}
\end{figure}

In Fig.\ref{fig5} we find an oscillatory feature at around
$10^{-7}$ Hz, which corresponds to the mode entering the horizon at the QGP
phase transition. At this energy scale, $\sim 180$ MeV, the effective
number of relativistic species changes drastically, giving a sharp feature and oscillation in $\Omega_h$. To understand this, let us consider the simple analytical model employed in subsection \ref{gstar2}. Eq.(\ref{3.2.6d}) is the mode which experienced such a change of $g_*$ and its coefficients $A, B, C$, and $D$ are 
\begin{eqnarray}
\label{A.1.1}
& &A(s, n)=\\
&\!&\frac{\pi}{4s^{3/2}}
\big[-2s Y_{1+\sqrt{1+4n}/2}(s)\sin{s}
+Y_{\sqrt{1+4n}/2}(s)\left(-2s\cos{s}+(3+\sqrt{1+4n})\sin{s}\right)\big],\nonumber
\end{eqnarray}
\begin{eqnarray}
\label{A.1.2}
& &B(s, n)=\\
&\!&-\frac{\pi}{4s^{3/2}}
\big[-2s J_{1+\sqrt{1+4n}/2}(s)\sin{s}
+J_{\sqrt{1+4n}/2}(s)\left(-2s\cos{s}+(3+\sqrt{1+4n})\sin{s}\right)\big],\nonumber
\end{eqnarray}
\begin{eqnarray}
\label{A.1.3}
&\!&C(s, t, n)=\nonumber\\
&\!&\frac{\pi}{4\sqrt{st}}\sec{n\pi}
\big[-2t J_{-n-3/2}(t)\cos{t}
\left(J _{n+1/2}(s)(s\cos{s}-\sin{s})+sJ _{n+3/2}(s)\sin{s}\right)\nonumber\\
&\!&-2J _{-n-1/2}(t)
\left(J _{n+1/2}(s)(s\cos{s}-\sin{s})+sJ _{n+3/2}(s)\sin{s} \right)(\cos{t}+t\sin{t})\\
&\!&+2\left(J _{-n-1/2}(s)(s\cos{s}-\sin{s})-sJ _{-n-3/2}(s)\sin{s}\right)\nonumber\\
&\!&\left(-tJ _{n+3/2}(t)\cos{t}+J _{n+1/2}(t)(\cos{t}+t\sin{t})\right)\big],\nonumber
\end{eqnarray}
\begin{eqnarray}
\label{A.1.4}
&\!&D(s, t, n)=\nonumber\\
&\!&\frac{\pi}{4\sqrt{st}}\sec{n\pi}
\big[2t J_{n+1/2}(t)\cos{t}
\left(J _{-n-1/2}(s)(-s\cos{s}+\sin{s})+sJ _{-n-3/2}(s)\sin{s}\right)\nonumber\\
&\!&-2J _{-n-1/2}(t)
\left(J _{n+1/2}(s)(s\cos{s}-\sin{s})+sJ _{n+3/2}(s)\sin{s} \right)(t\cos{t}-\sin{t})\\
&\!&-2\sin{t}\big(tJ _{-n-3/2}(t)
\left(J _{n+1/2}(s)(s\cos{s}-\sin{s})+sJ _{n+3/2}(s)\sin{s}\right)\nonumber\\
&\!&+\left(-tJ _{n+3/2}(t)\cos{t}+J _{n+1/2}(t)\right)
\left(sJ_{-n-3/2}(s)\sin{s}+J_{-n-1/2}(s)(-s\cos{s}+\sin{s})\right)\big)\big],\nonumber
\end{eqnarray}
where $s\equiv k\tau_2$, $t\equiv k\tau_1$ and $s\le t$. Here, $J_n(x)$
and $Y_n(x)$ are the Bessel function and Neumann function, respectively.
At this time, we are interested in different $k$-modes,
$k_1<k_2$. (However we evaluate $\Omega_h(k)$ at the same time, $\tau$.)
We obtain 
\begin{eqnarray}
\label{A.1.5}
\frac{\Omega^{\rm{in}}(k_2)}{\Omega^{\rm{out}}(k_1)}
&\simeq& \frac{k_2^2 h^2(k_2\tau)}{k_1^2 h^2(k_1\tau)},\nonumber\\
&=&\left(\frac{k_2}{k_1}\right)^2\left(\frac{\tau_2}{\tau_1}\right)^{2n}\left[C(k_2)\frac{j_0(k_2\tau)}{j_0(k_1\tau)}+D(k_2)\frac{y_0(k_2\tau)}{j_0(k_1\tau)}\right]^2,\nonumber\\
&\approx&\left(\frac{k_2}{k_1}\right)^2\left(\frac{\tau_2}{\tau_1}\right)^{2n}\left[C(k_2)\frac{k_1}{k_2}+D(k_2)\frac{k_1}{k_2}\right]^2
=\left(\frac{\tau_2}{\tau_1}\right)^{2n}\left[C(k_2)+D(k_2)\right]^2,\\
&\to&\left(\frac{\tau_2}{\tau_1}\right)^{2n}=\left(\frac{s}{t}\right)^{2n},
\end{eqnarray}
where $\simeq$ denotes the subhorizon limit, $\approx$ denotes the asymptotic limit as $k\tau\to$large, and $\to$ denotes the limit in $\Delta k\equiv k_2-k_1 \to 0$. Eq.~(\ref{A.1.5}) tells us the exact ratio between different $k$-modes. 
While we obtained only the suppression factor, $\left(\tau_2/\tau_1\right)^{2n}$, in subsection~\ref{gstar2}, we now also obtain the oscillatory factor, $[C+D]^2$. Fig.~\ref{CplusD2} shows that the factor, $[C+D]^2$, oscillates and takes on unity at $\Delta \tau \to0$ regardless of $n$. Here, $n=5$ represents $g_*\propto \tau^{-30}$, which is an extremely drastic change. 
This gives us a complete analytical account of the shape of Fig.~\ref{fig5}.

\begin{acknowledgments}
We are grateful to Joshua Adams for carefully reading and commenting on an early version of the manuscript. EK acknowledges support from the Alfred P. Sloan Foundation.
\end{acknowledgments}


\end{document}